\documentclass{iopart}
\usepackage{setstack,iopams}
\usepackage{psfig,cite}

\begin{document}

\title[Transition from regular to irregular motions]{The Transition from Regular to Irregular Motions, Explained as Travel
on Riemann Surfaces}

\author{F Calogero$^{1,2}$, D G\'omez-Ullate$^{3}$, P M Santini$^{1,2}$, and M Sommacal$^4$}

\address{$^1$\ Dipartimento di Fisica, Universit\`{a} di Roma ``La Sapienza'', Roma, Italy. }

\address{$^2$\ Istituto Nazionale di Fisica Nucleare, Sezione di Roma, Italy. }

\address{$^3$\ Dep. Matem\`atica Aplicada I, Universitat Polit\`ecnica de Catalunya, $\,\,$ ETSEIB, Av. Diagonal 647, 08028 Barcelona,
Spain.}

\address{$^3$\ SISSA, Triste, Italy.}

\eads{\mailto{francesco.calogero@roma1.infn.it},
  \mailto{david.gomez-ullate@upc.edu}, \mailto{paolo.santini@roma1.infn.it}, \mailto{sommacal@sissa.it}}
\begin{abstract}
We introduce and discuss a simple Hamiltonian dynamical system,
interpretable as a $3$-body problem in the (\textit{complex})
plane and providing the prototype of a mechanism explaining the
transition from \textit{regular }to \textit{irregular }motions as
travel on Riemann
surfaces. The interest of this phenomenology -- illustrating the \textit{%
onset} in a \textit{deterministic} context of \textit{irregular
}motions -- is underlined by its generality, suggesting its
eventual relevance to understand natural phenomena and
experimental investigations. Here only some of our main findings
are reported, without detailing their proofs: a more complete
presentation will be published elsewhere.
\end{abstract}

 \maketitle

\section{Introduction}

Purpose and scope of this paper is to introduce and discuss a simple
Hamiltonian dynamical system describing the motion of $3$ particles in the (%
\textit{complex}) plane. This \mbox{$3$-body} problem is the
prototype of a class of models \cite{1,CSCFS,C2} that feature a
transition from \textit{very simple\ }(even \textit{isochronous})
to \textit{quite complicated\ }motions characterized by a
\textit{sensitive dependence} both on the initial data and the
parameters (``coupling constants'') of the model. This transition
can be explained as travel on Riemann surfaces. The interest
of this phenomenology -- illustrating the \textit{onset} in a \textit{%
deterministic} context of \textit{irregular }motions -- is
underlined by its generality \cite{1,C2}, suggesting its eventual
relevance to understand natural phenomena and experimental
investigations. The novelty of the model treated herein is that it
allows a quite explicit mathematical treatment. Here only some of
our main findings are reported, without detailing their proofs: a
more complete presentation will be published elsewhere
\cite{CGSS}.

The idea that the \textit{integrable} or \textit{nonintegrable}
character of a dynamical system is closely related to the
\textit{analytic} structure of its solutions as functions of the
independent variable (``time'', but considered as a
\textit{complex} variable) goes back to such eminent
mathematicians as Carl Jacobi, Sophia Kowalewskaya, Henri
Poincar\'{e}, Paul Painlev\'{e} and his school. Some of us heard
illuminating discussions of this notion by Martin Kruskal, whose
main ideas -- a synthetic if overly terse rendition of which might
be the statement that \textit{integrability is compatible with the
presence of multivaluedness but only provided this is not
excessive} -- can be gleaned from some papers written by himself
and some of his collaborators \cite{Kru}, or by others who
performed theoretical and numerical investigations motivated by
his ideas \cite{BouEtAl}. The results presented below constitute
progress along this line of thinking. For a more detailed analysis
we refer the interested reader to the papers where more complete
versions are presented of our findings \cite{CGSS}.

\bigskip

\section{Results}

The model we introduce and discuss in this paper is characterized by the
following equations of motion:%
\begin{equation}
\dot{z}_{n}=-\rmi\,\omega \,z_{n}+\frac{g_{n+2}}{z_{n}-z_{n+1}}+\frac{g_{n+1}}{%
z_{n}-z_{n+2}}~.  \label{EqMot}
\end{equation}%
\textit{Notation}: here and hereafter indices such as $n,$ $m$
range from $1$ to $3$ and are defined $ \mbox{mod}(3);$
superimposed dots indicate differentiations with respect to the
\textit{real} independent time variable
$t$; the dependent variables $z_{n}\equiv z_{n}(t)$\textit{\ }are \textit{%
complex}, and indicate the positions of $3$ point ``particles''
moving in the \textit{complex }$z$-plane; $\rmi\equiv \sqrt{-1}$
is the \textit{imaginary} unit; the parameter $\omega $ is
\textit{positive}, and it sets the time
scale via the basic period%
\begin{equation}
T=\frac{\pi }{\omega }~;  \label{T}
\end{equation}%
the $3$ quantities $g_{n}$ are arbitrary coupling constants, but in this
paper we restrict consideration to the case in which they are all \textit{%
real} and moreover satisfy the ``semisymmetrical'' restriction
\begin{equation}
g_{1}=g_{2}=g~,~~~g_{3}=f~,  \label{Symm}
\end{equation}%
entailing that the two particles with labels $1$ and $2$ are
\textit{equal}, while particle $3$ is \textit{different}. More
special cases are the ``fully
symmetrical'', or``integrable'', one characterized by the equality of \textit{%
all }$3$ coupling constants,%
\begin{equation}
f=g~,~~~g_{1}=g_{2}=g_{3}=g~,  \label{Integr}
\end{equation}%
and the ``two-body'' one, with only one nonvanishing coupling
constant,
\numparts
\begin{equation}
g_{1}=g_{2}=g=0,~g_{3}=f\neq 0~.  \label{TwoBodya}
\end{equation}%
In this latter case clearly
\begin{equation}
z_{3}(t)=z_{3}(0)\,\exp (-\rmi\,\omega \,t)  \label{TwoBodyb}
\end{equation}%
and the remaining \textit{two-body} problem is easily solvable,%
\begin{eqnarray}
\fl \quad z_{s}(t) =\exp (-\rmi\,\omega \,t)\,\Bigg[ \frac{1}{2}\
[ z_{1}(0)+z_{2}(0)]   \nonumber \\ \fl \qquad\qquad\quad
-(-)^{s}\,\Big\{ \frac{1}{4}\ [ z_{1}(0)-z_{2}(0)]
^{2}+f\ \frac{\exp (2\,\rmi\,\omega \,t)-1}{2\,\rmi\,\omega }\Big\} ^{\,1/2}%
\Bigg] ~, \quad s =1,2~.\label{TwoBodyc}
\end{eqnarray}%
\endnumparts
The justification for labelling the fully symmetrical case
(\ref{Integr}) as ``integrable'' will be clear from the following
(or see Section 2.3.4.1\ of \cite{C2}). The treatment of the more
general case with $3$ different coupling constants $g_{n}$ is
outlined in \cite{CGSS}.

Note that the equations of motion (\ref{EqMot}) are of
``Archimedian'', rather than ``Newtonian'', type, inasmuch as they
imply that the ``velocities'', rather than the ``accelerations'',
are determined by the ``forces''. These equations of motion are
Hamiltonian, indeed they follow in the standard manner from the
Hamiltonian function
\begin{equation}
H({\bf z},{\bf p})=\sum\limits_{n=1}^{3}\left[ -\rmi\,\omega
\,z_{n}\,p_{n}+g_{n}\,\frac{p_{n+1}-p_{n+2}}{z_{n+1}-z_{n+2}}\right]
~. \label{H}
\end{equation}%
And they can be reformulated \cite{CGSS} as, still Hamiltonian,
\textit{real} (and \textit{covariant}, even
\textit{rotation-invariant}) equations describing the motion of
three point particles in the (\textit{real}) horizontal plane.

The following \textit{qualitative} analysis (confirmed by our \textit{%
quantitative} findings, see below) is useful to get a first idea
of the nature of the motions entailed by our model. For
\textit{large} values of (the modulus of) $z_{n}$ the ``two-body
forces'' represented by the last two terms in the right-hand side
of (\ref{EqMot}) become \textit{negligible} with respect to the
``one-body (linear) force'' represented by the first term,
hence in this regime $\dot{z}_{n}\approx -i\,\omega \,z_{n}$ entailing $%
z_{n}(t)\approx const\,\exp (-i\,\omega \,t).$ One thereby infers
that, when a particle strays far away from the origin in the
complex $z$-plane, it tends to rotate (clockwise, with period
$2\,T$) on a circle: hence the
first \textit{qualitative} conclusion that \textit{all motions are confined}%
. Secondly, the \textit{two-body} forces cause a \textit{singularity}
whenever there is a \textit{collision} of \textit{two} (or all \textit{three}%
) of the particles, and become dominant whenever \textit{two} particles get
very close to each other, namely in the case of \textit{near misses}. But if
the three particles move \textit{aperiodically} in a \textit{confined%
} region (near the origin) of the \textit{complex} $z$-plane, an \textit{%
infinity} of \textit{near misses} shall indeed occur. And since
the outcome of a \textit{near miss} is generally quite {\em
different} (whenever the two particles involved in it are {\em
different}) depending on which side the particles slide past each
other -- and this, especially in the case of \textit{very close}
near misses, depends \textit{sensitively} on the initial data of
the trajectories under consideration -- we see here a mechanism
causing a \textit{sensitive dependence} of the time evolution on
its initial data. This suggests that our model (\ref{EqMot}), in
spite of its simplicity, might also support quite complicated
motions, possibly even
displaying an ``unpredictable'' evolution in spite of its \textit{deterministic%
} character. This hunch is confirmed by the results reported below.

To investigate the dynamics of our ``physical'' model
(\ref{EqMot}) it is convenient to introduce an ``auxiliary''
model, obtained from it via the
following change of dependent and independent variables:%
\begin{equation}
z_{n}(t)=\exp \left( -\rmi\,\omega \,t\right) \,\zeta _{n}\left(
\tau \right) ~,~~~\tau \left( t\right) =\frac{\exp
(2\,\rmi\,\omega \,t)-1}{2\,\rmi\,\omega }~. \label{zita}
\end{equation}%
Note that initially the coordinates $z_{n}$ and $\zeta _{n}$
coincide:%
\begin{equation}
z_{n}(0)=\zeta _{n}(0)~.  \label{zitazero}
\end{equation}%
The equations of motion of the auxiliary model follow immediately from (\ref%
{EqMot}) via (\ref{zita}) (or, even more directly, by noting that, for $%
\omega =0,$ $\tau =t$ and $z_{n}(t)=\zeta _{n}(\tau ))$:%
\begin{equation}
\zeta _{n}^{\prime }=\frac{g_{n+2}}{\zeta _{n}-\zeta _{n+1}}+\frac{g_{n+1}}{%
\zeta _{n}-\zeta _{n+2}}~.  \label{EqZita}
\end{equation}%
Here of course the appended prime denotes differentiation with respect to
the (\textit{complex}) variable $\tau $.

The definition of $\tau \left( t\right)$  implies that as the
(\textit{real}) time variable $t$ evolves onwards from $t=0,$ the
\textit{complex }variable $\tau $ travels round and round, making
a full tour (counterclockwise) in every time interval $T,$ on the
circle $C$ the diameter of which, of length $d=1\,/\,\omega ,$
lies on the imaginary axis in the \textit{complex }$\tau $-plane,
with one end at the origin, $\tau =0,$ and the other at $\tau
=\rmi\,/\,\omega $ (draw this circle!). Hence these relations,
(\ref{zita}), entail that if $\zeta _{n}\left( \tau \right) $ is
\textit{holomorphic} as a function of the \textit{complex}
variable $\tau $ in the closed disk $D$ encircled by the circle
$C$, the corresponding function $z_{n}\left( t\right) $ is
\textit{periodic} in the \textit{real}
variable $t$ with period $2\,T$ (indeed \textit{antiperiodic }with period $%
T) $:%
\begin{equation}
z_{n}(t+T)=-z_{n}(t)~,~~~z_{n}(t+2\,T)=z_{n}(t)~.  \label{Periodic}
\end{equation}%
But it is easy to prove \cite{CGSS} that the solution $\zeta _{n}\left( \tau
\right) $ of (\ref{EqZita}) is \textit{holomorphic} (at least) in the
circular disk $D_{0}$ centered at the origin of the \textit{complex} $\tau $%
-plane and having the radius%
\begin{equation}
r=\frac{\left( \underset{n,m=1,2,3;\,m\neq n}{\min }\left\vert \zeta
_{n}(0)-\zeta _{m}(0)\right\vert \right) ^{\,2}}{128\,\underset{n=1,2,3}{%
\max }\left\vert g_{n}\right\vert }~.  \label{dzero}
\end{equation}%
One may therefore conclude that our physical system (\ref{EqMot}) is \textit{%
isochronous with period }$2\,T,$ see (\ref{T}). Indeed an \textit{%
isochronous }system is characterized by the property to possess one or more
\textit{open} sectors of its phase space, each having of course \textit{full
dimensionality}, such that \textit{all} motions in each of them are \textit{%
completely periodic} with the same \textit{fixed period} (the periods may be
different in these different sectors of phase space, but must be fixed, i e.
independent of the initial data, within each of these sectors): and in our
case clearly (at least) \textit{all} the motions characterized by initial
data $z_{n}\left( 0\right) $ such that%
\begin{equation}
\underset{n,m=1,2,3;\\~m\neq n}{\min }\left\vert
z_{n}(0)-z_{m}(0)\right\vert
>16\,\sqrt{\frac{\underset{n=1,2,3}{\max }\left\vert g_{n}\right\vert }{%
2\,\omega }}  \label{CondIso}
\end{equation}%
are \textit{completely periodic} with period $2\,T$, see (\ref{Periodic}),
since this inequality, implying (via (\ref{zitazero}) and (\ref{dzero})) $%
r>d,$ entails that $\zeta _{n}(\tau )$ is \textit{holomorphic} (at least) in
a disk $D_{0}$ that includes, in the \textit{complex }$\tau $-plane, the
disk $D$.

This argument is a first demonstration of the usefulness of the ``trick'' (\ref%
{zita}), associating the auxiliary system (\ref{EqZita}) to our physical
system (\ref{EqMot}). More generally, this relationship (\ref{zita}) allows
to infer the main characteristics of the \textit{time evolution} of the
solutions $z_{n}\left( t\right) $ of our physical system (\ref{EqMot}) from
the \textit{analyticity properties} of the corresponding solutions $\zeta
_{n}\left( \tau \right) $ of the auxiliary system (\ref{EqZita}): indeed the
evolution of $z_{n}(t)$ as the time $t$ increases from the initial value $%
t=0 $ is generally related via (\ref{zita}) to the values taken by $\zeta
_{n}\left( \tau \right) $ when $\tau $ rotates (counterclockwise, with
period $T)$ on the circle $C$ in the \textit{complex }$\tau $-plane and
correspondingly $\zeta _{n}(\tau )$ travels on the Riemann surface
associated to its \textit{analytic} structure as a function of the \textit{%
complex} variable $\tau $. Suppose for instance that the
\textit{only} singularities of $\zeta _{n}(\tau )$ in the
\textit{finite} part of the \textit{complex} $\tau $-plane are
\textit{square-root branch points}, as it is indeed the case for
our model (\ref{EqMot}) at least for a range of values of the
ratio of the coupling constants $f$ and $g,$ see (\ref{Symm}) and
below.\footnote{The nature of these singularities is generally
independent from the particular solution under consideration and
can be
easily ascertained \cite{CGSS} via local analyses \textit{\`{a} la Painlev%
\'{e}} of the generic solution of the equations of motion (\ref{EqZita}),
while the number and especially the locations of these singularities
 depend on the specific solution
under consideration and their identification requires a more
detailed
knowledge than can be obtained by a local analysis \textit{\`{a} la Painlev%
\'{e}}.} Then the \textit{isochronous} regime corresponds to
initial data such that the corresponding solution $\zeta
_{n}\left( \tau \right) $ has \textit{no branch points} inside the
circle $C$ on the main sheet of its Riemann surface (i. e. that
characterized by the initial data). Moreover, if there is a
\textit{finite} (\textit{nonvanishing}) number of branch points
inside the circle $C$ on the main sheet of the Riemann surface of
$\zeta _{n}\left( \tau \right) ,$ and a \textit{finite} number of
branch points inside the circle $C$ on all the sheets that are
accessed by traveling on the Riemann surface round and round on
the circle $C$, then clearly the corresponding solution $z_{n}(t)$
is still a \textit{completely periodic} function of the time $t,$
but now its period is a \textit{finite integer multiple} $jT$ of
the basic period $T,$ the value of $j$ depending of course on the
number of sheets that get visited along this travel before
returning
to the main sheet. Hence, in particular, whenever the \textit{total} number $%
q$ of (\textit{square-root}) branch points of the solution $\zeta
_{n}(\tau ) $ of the auxiliary problem (\ref{EqZita}) is
\textit{finite}, the corresponding solution $z_{n}(t)$ of our
physical model (\ref{EqMot}) is \textit{completely periodic},
although possibly with a \textit{very large} period (if $q$ is
\textit{very large}) the value of which may depend, possibly quite
sensitively, on the initial data. On the other hand if the number
of (\textit{square-root}) branch points possessed by the generic
solution $\zeta _{n}(\tau )$ is \textit{infinite}, and the Riemann
surface associated with the function $\zeta _{n}(\tau )$ has an
\textit{infinite} number of sheets (as it can happen in our case,
see below), then it is possible that, as $\tau $ goes round and
round on the circle $C,$ the corresponding value of $\zeta
_{n}\left( \tau \right) $ travels on this Riemann surface without
ever returning to its main sheet, entailing that the time
evolution of the corresponding function\textit{\ }$z_{n}(t)$ is
\textit{aperiodic}, and that it depends \textit{sensitively} on
the initial data inasmuch as these data characterize the positions
of the branch points hence the structure of the Riemann surface.

This terse analysis entails an important distinction among all these (%
\textit{square-root})\textit{\ }branch points: the ``active''
branch-points are those located \textit{inside} the circle $C$ on
sheets of the Riemann surface accessed -- when starting from the
main sheet -- by traveling round and round on that circle, so that
they do affect the subsequent sequence of sheets that get visited;
while the ``inactive'' branch points are, of course, those that
fall \textit{outside} the circle $C,$ as well as those that are
located \textit{inside} the circle $C$ but on sheets of the
Riemann surface that do not get visited while traveling round and
round on that circle (starting from the main sheet) and that
therefore do \textit{not} influence
the time-evolution of the corresponding solution of our physical system (\ref%
{EqMot}). This distinction is of course influenced by the initial
data of the problem, that characterize the initial pattern of
branch points; clearly it is not just a ``local'' characteristic
of each branch point depending only on its position (for instance,
\textit{inside} or \textit{outside} the circle $C)$: it depends on
the overall structure of the Riemann surface, for instance if
there is no branch point on its main sheet -- that containing the
point of departure of the travel round and round on the circle $C$
-- then clearly \textit{all} the other branch points are
\textit{inactive}, irrespective of their location.

\begin{figure}[htbp]
\begin{center}
    \noindent\psfig{figure=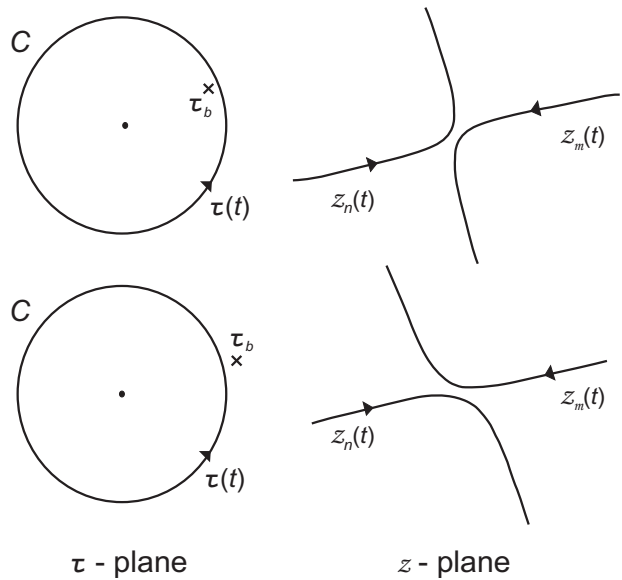,width=3.2in}
    \caption{ Scattering of two bodies in the three-body problem
    \eref{EqMot} corresponding to a {\em near miss}. The two outcomes originate from two sets of
    initial
    data close to each other
   such that a square-root branch point $\tau_b$ falls on
    different sides of the  circle $C$.}
    \label{fig1}
    \end{center}
\end{figure}

Let us also emphasize that, whenever an \textit{active} branch
point is \textit{quite close} to the circle $C,$ it corresponds to
a \textit{near miss }involving \textit{two} particles of our
physical model (\ref{EqMot}), at which these two particles scatter
against each other almost at right angles (corresponding to the
\textit{square-root} nature of the branch point). The difference
between the cases in which such a branch point falls \textit{just
inside} respectively \textit{just outside} the circle $C$
corresponds to a \textit{near miss} in which the two particles
slide past each other on one side respectively on the other (see
figure \ref{fig1}), and this makes a substantial difference as
regards the subsequent evolution of our $3$-body system (unless
the two particles are equal).
 The closer the \textit{near miss%
}, the more significant this effect is, and the more \textit{sensitive} it
is on the \textit{initial data}, a tiny change of which can move the
relevant branch point from one side to the other of the circumference of the
circle $C$ and correspondingly drastically affect the outcome of the \textit{%
near miss}. This is the mechanism that accounts for the fact that, when the
initial data are in certain sectors of their phase space (of course quite
different from that characterized by the inequalities (\ref{CondIso})), the
resulting motion of the physical $3$-body problem (\ref{EqMot}) is \textit{%
aperiodic}, indeed nontrivially so: in such cases (as we show below) the
\textit{aperiodicity} is indeed associated with the coming into play of an
\textit{infinite} number of (\textit{square-root}) branch points of the
corresponding solution of the auxiliary problem (\ref{EqZita}) and\textit{\ }%
correspondingly with an \textit{infinite }number of \textit{near} \textit{%
misses} experienced by the particles throughout their time evolution, this
phenomenology being clearly characterized by a \textit{sensitive dependence}
on the initial data.

This mechanism to explain the transition from \textit{regular }to \textit{%
irregular} motions -- and in particular from an \textit{isochronous} regime
to one featuring \textit{unpredictable }aspects -- was already discussed
\cite{CSCFS} in the context of certain many-body models somewhat analogous
to that studied herein. But those treatments were limited to providing a
\textit{qualitative} analysis such as that presented above and to
ascertaining its congruence with \textit{numerical solutions }of these
models. The interest of the simpler model introduced and discussed herein is
to allow a detailed, \textit{quantitative} understanding of this
phenomenology. This is based on the following explicit solution of our model
(\ref{EqMot}), obtained \cite{CGSS} via the auxiliary problem (\ref{EqZita}%
): \numparts\label{prueba}
\begin{eqnarray}\fl\quad
z_{s}(t)=Z\,\exp \left( -\rmi\,\omega \,t\right) -\frac{1}{2}\left( \frac{f+8\,g%
}{6\,\rmi\,\omega }\right) ^{1/2}\,\left[ 1+\eta \,\exp \left(
-2\,\rmi\,\omega
\,t\right) \right] ^{\,1/2}\cdot &&  \nonumber \\
\quad \cdot \left\{ \left[ \check{w}\left( t\right) \right]
^{\,1/2}-\left( -\right) ^{s}\,\left[ 12\,\mu -3\,\check{w}\left(
t\right) \right] ^{\,1/2}\right\} ~,\qquad\quad s=1,2~, &&
\label{ZSolution}\\ \label{ZSolutionb}
\fl \quad z_{3}\left( t\right) =Z\,\exp \left( -\rmi\,\omega \,t\right) +\left( \frac{%
f+8\,g}{6\,\rmi\,\omega }\right) ^{\,1/2}\,\left[ 1+\eta \,\exp
\left(
-2\,\rmi\,\omega \,t\right) \right] ^{\,1/2}\,\left[ \check{w}\left( t\right) %
\right] ^{\,1/2}~.
\end{eqnarray}%
\endnumparts
Here the function $\check{w}(t)$ is defined via the relation
\begin{equation}
\check{w}(t)=w\left[ \xi (t)\right],  \label{WtildeW}
\end{equation}%
with%
\begin{equation}
\xi (t)=R\,\left[ \eta +\exp \left( 2\,\rmi\,\omega \,t\right) \right] =\bar{\xi%
}+R\,\exp \left( 2\,\rmi\,\omega \,t\right) ~,  \label{KSI}
\end{equation}%
and $w(\xi )$ implicitly defined by the \textit{nondifferential} equation
\begin{equation}
\left( w-1\right) ^{\,\mu -1}\,w^{\,-\mu }=\xi ~.  \label{EqWtilde}
\end{equation}%
The parameter $\mu $ is defined in terms of the coupling constants $g$ and $%
f,$ see (\ref{Symm}), as follows:%
\begin{equation}
\mu =\frac{f+2\,g}{f+8\,g}~,  \label{mu}
\end{equation}%
and in (13)-(\ref{KSI}) the \textit{three }constants $Z,$ $R,$ and $\eta $ (or $%
\bar{\xi})$ are defined in terms of the $3$\textit{\ }initial data
$z_{n}(0)$ as follows: \numparts
\begin{eqnarray}\fl\qquad\quad\quad
Z=\frac{z_{1}(0)+z_{2}(0)+z_{3}(0)}{3}~,  \label{Para}\\[4pt]
\fl\qquad\quad\quad R=\frac{3\,\left( f+8\,g\right) }{2\,\rmi\,\omega \,\left[ 2%
\,z_{3}(0)-z_{1}(0)-z_{2}(0)\right] ^{\,2}}\,\left[ 1-\frac{1}{\check{w}(0)}%
\right] ^{\,\mu -1}~,  \label{Parb}\\[4pt]
\fl\qquad\quad\quad \bar{\xi}=R\,\eta ~,  \label{Parc}\\[4pt]
\fl\qquad\quad\quad\eta =\frac{\rmi\,\omega \,\left\{ \left[
z_{1}(0)-z_{2}(0)\right] ^{\,2}+\left[ z_{2}(0)-z_{3}(0)\right]
^{\,2}+\left[ z_{3}(0)-z_{1}(0)\right] ^{\,2}\right\} }{3\,\left(
f+2\,g\right) }-1~,  \label{Pard}\\[4pt]
\fl\qquad\quad\quad\check{w}(0)=\frac{2\,\mu \,\left[
2\,z_{3}(0)-z_{1}(0)-z_{2}(0)\right]
^{\,2}}{\left[ z_{1}(0)-z_{2}(0)\right] ^{\,2}+\left[ z_{2}(0)-z_{3}(0)%
\right] ^{\,2}+\left[ z_{3}(0)-z_{1}(0)\right] ^{\,2}}~.
\label{Pare}
\end{eqnarray}
\endnumparts

Note that the constant $Z$ is the initial value of the center of
mass of the system, and indeed the first term in the right-hand
side of the solution (13) represents the motion of the center of
mass of the system: just a circular motion around the origin, with
a constant velocity entailing a period $2\,T$. Since the rest of
the motion is independent of the behavior of the center of mass,
in the study of this model attention can be restricted without
significant loss of generality to the case when the center of mass
does not move, $Z=0.$

The nontrivial aspects of the motion are encoded in the time
evolution of the function $\check{w}(t)$, see (13) and
(\ref{WtildeW}): let us emphasize in this connection that the
dependent variable $w(\xi )$ is that solution of the
\textit{nondifferential }equation (\ref{EqWtilde}) uniquely
identified by continuity, as the time $t$ unfolds, hence as the
variable $\xi \equiv \xi \left( t\right) $ goes round and round,
in the \textit{complex }$\xi $-plane, on the circle $\Xi $ with
center $\bar{\xi}$ and radius $\left\vert R\right\vert $ (see
(\ref{KSI})), from the initial datum assigned at $t=0$,
\begin{equation}
w\left[ \xi \left( 0\right) \right] =w\left( \bar{\xi}+R\right) =\check{w}%
\left( 0\right) ~,  \label{win}
\end{equation}%
see (\ref{Pare}). This specification of the initial value $\check{w}(0)$ is
relevant, because generally the \textit{nondifferential} equation (\ref%
{EqWtilde}) has more than a single solution, in fact possibly an \textit{%
infinity} of solutions, see below.

It is clear from (13) that the time evolution of the solution $%
z_{n}(t)$ of our model (\ref{EqMot}) is \textit{mainly} determined
by the time evolution of the function $\check{w}(t)$. Indeed,
\begin{enumerate}
\item The factor $\left[ \eta \,\exp \left( -2\,\rmi\,\omega
\,t\right) -1\right] ^{\,1/2}$ displays a
quite simple time evolution, \textit{periodic} with period $T$ if $%
\left\vert \eta \right\vert <1$ and \textit{antiperiodic} with period $T$
hence \textit{periodic} with period $2\,T$ if $\left\vert \eta \right\vert
>1 $.
\item If $\check{w}(t)$ is \textit{periodic} with period
$\check{T}$, its square root $\left[ \check{w}(t)\right]
^{\,1/2},$ appearing in the right-hand side of the solution
formulas (13), is clearly as well \textit{periodic} with period
$\check{T}$ or \textit{antiperiodic} with period $\check{T}$ hence
\textit{periodic} with period $2\,\check{T}$
depending whether the closed trajectory of $\check{w}(t)$ in the complex $%
\check{w}$-plane does not or does enclose the (branch) point
$\check{w}=0.$ \item Likewise the square root $\left[ 12\,\mu -3\,\check{w}\left( t\right) %
\right] ^{\,1/2}$ (see (\ref{ZSolution})) is also
\textit{periodic} with period $\check{T}$ or \textit{antiperiodic}
with period $\check{T}$ hence \textit{periodic} with period
$2\,\check{T}$ depending whether the closed trajectory of
$\check{w}(t)$ in the complex $\check{w}$-plane does not or does
enclose the (branch) point $\check{w}=4\,\mu $ (but note that a
change of sign of this square root only entails an exchange
between the two equal particles $1$ and $2$). \item In conclusion
one sees that -- provided one considers particles $1$ and $2$ as
\textit{indistinguishable} -- then, if
the time evolution of $\check{w}(t)$ is \textit{periodic} with period $%
\check{T}$, $\check{w}(t+\check{T})=\check{w}(t),$ the physical motion of
the $3$ particles $z_{n}(t)$ is also \textit{completely periodic} either
with the same period $\check{T}$ or with period $2\,\check{T},$ provided $%
\check{T}$ is an \textit{integer multiple} of $T$; \item Finally,
if the motion of $\check{w}(t)$ is not periodic then clearly the
functions $z_n(t)$ are also not periodic.
\end{enumerate}

Hereafter we only discuss the time evolution of the function
$\check{w}(t)$; actually, as explained below, in this paper we
limit our consideration to discussing the motion of a
\textit{generic} solution $\tilde{w}(t)=w\left[ \xi (t)\right] $
of the \textit{nondifferential} equation (\ref{EqWtilde}).
Moreover, we consider only {\em generic} solutions of
(\ref{EqMot}), namely those characterized by initial data that
exclude one of the following special outcomes:
\begin{enumerate}
\item[(a)] $\check{w}\left( t\right) $
takes, at some (\textit{real}) time $t_{a}$, the value $\check{w}%
(t_{a})=4\,\mu $ entailing a \textit{pair collision} of the $2$
equal particles occurring at this time,
$z_{1}(t_{a})=z_{2}(t_{a})$.
\item[(b)] $\check{w}\left( t\right) $ takes, at some (\textit{real}%
) time $t_{b}$, the value $\check{w}(t_{b})=\mu $ entailing a \textit{pair
collision} of the different particle with one of the $2$ equal particles
occurring at this time, $z_{1}(t_{b})=z_{3}(t_{b})$ or $%
z_{2}(t_{b})=z_{3}(t_{b})$. \item[(c)] The constant $\eta $ has
unit modulus, $\left\vert \eta \right\vert =1$, i. e. $\eta =\exp
(2\,\rmi\,\omega \,t_{c})$ with $t_{c}$ \textit{real }(and of course defined $%
\mbox{mod}\,T$) which entails a \textit{triple collision} of the
$3$
particles occurring at the time $t_{c}$, $%
z_{1}(t_{c})=z_{2}(t_{c})=z_{3}(t_{c})$ .
\item[(c')] $\check{w}\left( t\right) $ vanishes at some (\textit{real}) time $t_{c}$, $%
\check{w}(t_{c})=0$ (but, as our notation suggests, this case (c') is just a
subcase of (c), although this is not immediately obvious from (\ref%
{ZSolution}) but requires using also (\ref{KSI}) and (\ref{EqWtilde})).
\end{enumerate}
The initial data that give rise to solutions having one of these
singularities form a set of null measure. It can be easily seen
that these singular solutions $z_{n}(t)$ of our physical problem
(\ref{EqMot}) correspond via (\ref{zita}) to special solutions
$\zeta _{n}\left( \tau \right) $ of our auxiliary problem
(\ref{EqZita}) possessing a \textit{branch point} that sits
\textit{exactly} on the circle $C$ in the complex $\tau $-plane:
more precisely,
\begin{enumerate}
\item[(a)] a \textit{square-root} branch point featured by $\zeta
_{1}(\tau )$ and $\zeta _{2}(\tau )$ but not by $\zeta _{3}(\tau
)$, \item[(b)] a \textit{square-root} branch point featured by all
$3$ functions $\zeta _{n}(\tau )$, \item[(c)] a branch point
featured by all $3$ functions $\zeta _{n}(\tau )$ the nature of
which depends on the parameter $\mu $.
\end{enumerate}
As mentioned above, in this paper we confine our treatment to discussing the
time evolution of a \textit{generic }root $\tilde{w}(t)$ of the \textit{%
nondifferential }equation (\ref{EqWtilde}) with (\ref{KSI}), and in
particular to identifying for which initial data its time evolution is
\textit{periodic}, and in such a case what the period is. Remarkably we find
out that, for (\textit{arbitrarily}) given initial data, \textit{all} these
roots have at most three different periods (one of which might be \textit{%
infinite}, signifying an \textit{aperiodic} motion); periods which
we are able to determine explicitly (although the relevant
formulas have some nontrivial, even ``chaotic", aspects, in a
sense that is made explicit
below). The question of identifying, among \textit{all} the roots $\tilde{w}%
_{j}(t)$ of this \textit{nondifferential }equation
(\ref{EqWtilde}), the ``physical'' one $\check{w}(t)$ i. e. the
one that evolves from the initial datum (\ref{Pare}), and in
particular of specifying the character of its time evolution among
the (at most $3$) alternatives discussed below, is a
technically demanding job the solution of which shall be reported in \cite%
{CGSS}. Let us re-emphasize that the time evolution of $\tilde{w}(t)\equiv w%
\left[ \xi (t)\right] $ coincides with the evolution of a \textit{generic}
root $w\left( \xi \right) $ of (\ref{EqWtilde}) as the independent variable $%
\xi $ travels (making a complete counterclockwise tour in the \textit{complex%
} $\xi $-plane in every time interval $T$) on the circle $\Xi $ with center $%
\bar{\xi}$ and radius $\left\vert R\right\vert $, see (\ref{KSI}),
and correspondingly the dependent variable $w(\xi )$ travels on
its Riemann surface. Note that this Riemann surface is completely
defined by the single parameter $\mu $, see (\ref{mu}) and
(\ref{EqWtilde}), while the circle $\Xi $ is defined by the
initial data of the problem, see (\ref{KSI}) with (18).

What therefore remains to be discussed is the \textit{analytic} structure of
the multivalued function $w\left( \xi \right) $ defined implicitly by the
\textit{nondifferential} equation (\ref{EqWtilde}) or, equivalently but more
directly, the time dependence of the corresponding function $\tilde{w}\left(
t\right) \equiv \tilde{w}\left[ \xi \left( t\right) \right] $. To begin
with, we consider the case in which the parameter $\mu $ is \textit{rational}%
,
\begin{equation}
\mu =\frac{p}{q}~,  \label{mupq}
\end{equation}%
with $p$ and $q$ coprime integers and $q$ \textit{positive}, $q>0$. The
extension of the results to the case of \textit{irrational} $\mu $ is made
subsequently; although, to avoid repetitions, we present below some results
in a manner already appropriate to include also the more general case with $%
\mu $ \textit{real}.

In the \textit{rational} case (\ref{mupq}) the
\textit{nondifferential} equation that determines the ``dependent
variable'' $\tilde{w}\left( t\right) $ in terms of the
``independent variable'' $t$ becomes \textit{polynomial}, and
takes one of the following $3$ forms depending on the value of the
parameter $\mu ,$ see (\ref{mupq}): \numparts
\begin{eqnarray}
\left( \tilde{w}-1\right) ^{\,p-q}=\left[ \bar{\xi}+R\,\exp \left(
2\,\rmi\,\omega \,t\right) \right] ^{\,q}\,\tilde{w}^{\,p},\qquad&\mbox{if }\mu >1,  \label{PolEqa}\\
\left[ \bar{\xi}+R\,\exp \left( 2\,\rmi\,\omega \,t\right) \right]
^{\,q}\,\left( \tilde{w}-1\right) ^{\,q-p}\,\tilde{w}^{\,p}=1,\qquad&\mbox{if }%
0<\mu <1,  \label{PolEqb}\\
\left[ \bar{\xi}+R\,\exp \left( 2\,\rmi\,\omega \,t\right) \right]
^{\,q}\,\left( \tilde{w}-1\right) ^{\,q+\left\vert p\right\vert }=\tilde{w}%
^{\,\left\vert p\right\vert },\qquad&\mbox{if }\mu <0.
\label{PolEqc}
\end{eqnarray}%
\endnumparts
The above expressions are polynomials (in the dependent variable
$\tilde{w}$) of degree $J$:
\begin{equation}\label{JEq}
J=\left\{%
\begin{array}{ll}
    p, & \hbox{if $\,\,\mu>1$;} \\
    q, & \hbox{if $\,\,0<\mu<1$;} \\
    q+|p|, & \hbox{if $\,\,\mu<0$.} \\
\end{array}%
\right.
\end{equation}
 As for the boundaries of these $3$
cases, let us recall that $\mu =1$ corresponds, via (\ref{mu}), to
$g=0,$ namely, see (\ref{Symm}), to the trivially solvable
\textit{two-body} case, see (5), while $\mu =0$ respectively $\mu
=\infty $ correspond, via (\ref{mu}), to $f+2\,g=0$ respectively
to $f+8\,g=0$ and require a
separate treatment, for which the interested reader is referred to \cite%
{CGSS}. Clearly the third case ($\mu <0$) becomes identical to the first ($%
\mu >1$) via the replacement
 \begin{equation}\fl\qquad \tilde{w}\mapsto
1-\tilde{w},\quad\bar{\xi} \mapsto -\bar{\xi},\quad R\mapsto
-R,\quad -p\mapsto p-q,\quad q-p\mapsto p \end{equation} without
modifying $q$; therefore in the following, without loss of
generality, we often forsake a separate discussion of this third
case.

Clearly the factor $\left[ \bar{\xi}+R\,\exp \left( 2\,\rmi\,\omega \,t\right) %
\right] ^{\,q},$ which carries all the time dependence in these polynomial
equations, is \textit{periodic} in $t$ with period $T$, see (\ref{T})
(except for the \textit{special} initial conditions entailing $\bar{\xi}=0,$
in which case this factor is instead periodic with the shorter period $%
T\,/\,q$) (for simplicity we continue to pursue our policy to
consider only the \textit{generic} case when this does
\textit{not} happen, referring the interested reader to
\cite{CGSS} for a more complete treatment).
 At issue is the
behaviour of the $J$ roots $\tilde{w}_{j}(t)$ of our polynomial
equation  (21) whose coefficients evolve in time periodically
with period $T$. Let us indicate with $\tilde{W}(t)\equiv \left\{ \tilde{w}%
_{j}(t);\,j=1,...,J\right\} $ the (\textit{unordered}) set of these $J$
roots. Obviously $\tilde{W}(t)$ is \textit{periodic} with period $T$, $%
\tilde{W}(t+T)=\tilde{W}(t)$: after one period $T$ the polynomial equation
is unchanged, hence the set of its $J$ roots is as well unchanged. But that
does \textit{not} imply that if one follows the time evolution of these $J$
roots, each of them will return to its own initial value after one period, $%
\tilde{w}_{j}(T)=\tilde{w}_{j}(0),$ $j=1,...,J.$ This outcome will indeed
obtain for the \textit{open} domain of initial data of our problem that
corresponds to the basic \textit{isochronous }regime, see (\ref{Periodic});
but it does not happen for other initial data, in which cases for instance a
\textit{generic} root, say $\tilde{w}_{j_{1}}(t)\equiv \tilde{w}(t),$ may
after one period land in the initial position of a different root, say $%
\tilde{w}(T)=\tilde{w}_{j_{2}}(0),$ and then after one more period end up in
the initial position of yet another root, $\tilde{w}(2\,T)=\tilde{w}%
_{j_{3}}(0),$ and so on. Eventually, of course, after a time $\tilde{T}=%
\tilde{j}\,T$ which is a \textit{finite integer} multiple $\tilde{j}$ of the
basic period $T$, with $1\leq \tilde{j}\leq J,$ the \textit{generic }root $%
\tilde{w}(t)$ shall necessarily return to its initial position, $\tilde{w}%
\left( \tilde{T}\right) =\tilde{w}\left( 0\right) ,$ entailing
that its evolution as a function of the time $t$ is
\textit{periodic} with this period $\tilde{T}$,
\begin{equation}
\tilde{w}\left( t+\tilde{T}\right) \equiv \tilde{w}\left( t+\tilde{j}%
\,T\right) =\tilde{w}\left( t\right) ~.
\end{equation}
This discussion clearly implies (via (\ref{ZSolution})) that, in
the case
now under consideration (with a \textit{rational} value of the parameter $%
\mu $, see (\ref{mupq})), \textit{all} solutions of our physical problem (%
\ref{EqMot}) with (\ref{Symm}) are \textit{completely periodic} with a
period which is either
\begin{equation}
\check{T}=\check{j}\,T~~~\mbox{with }1\leq \check{j}\leq J
\label{Ttilde}
\end{equation}%
and $J$ defined by (\ref{JEq}), or it is $2\,\check{T}$ (see the
discussion above following (\ref{win})). The remaining, crucial
question is: how does the value of the integer $\check{j}$ (which
might be \textit{quite large} if $J$ is \textit{quite large})
depend on the initial data of our problem? In this paper we
outline how to calculate, for given
initial data, \textit{all} the possible periods of the $J$ roots $\tilde{w}%
_{j}(t)$ of (21), and we display formulas providing (at most) $3$
alternative values for these periods; as already mentioned above,
the explanation of how to identify which one of these $3$ periods
corresponds to that of the ``physical'' root $\check{w}(t)$
entails a more detailed treatment which is reported in
\cite{CGSS}.

But before doing so let us emphasize that via this discussion the
time evolution of our original $3$-body problem -- describing the
time evolution of the three points $z_{n}(t)$ in the
\textit{complex }$z$-plane -- has been related to the time
evolution of the $J$ roots $\tilde{w}_{j}(t)$ of (21) in the
\textit{complex} $w$-plane, and in particular to the way they get
permuted among themselves over the time evolution after each
period $T$. As we will explain below, the possible complications
of the motions of our physical model (\ref{EqMot}) are
thereby~related to the mechanisms at play to permute these roots
among themselves when one watches their time evolution at
subsequent intervals $T,~2\,T,~3\,T$\thinspace and so on. And,
in this context, it is significant to note that, whenever $\mu $ is \textit{%
irrational}, one is in fact dealing with the dynamics of an \textit{infinite
}number of roots. This suggests that whenever the number $J$ of roots is
large, and even more so when $\mu $ is irrational (entailing $J=\infty $),
the time evolution of our physical model (\ref{EqMot}) might be quite
complicated, perhaps calling into play the theoretical tools of statistical
mechanics rather than those of few-body dynamics (but we postpone such
excursions to future publications).

As entailed by our discussion above, the issue of determining the value of
the integer $\tilde{j}$ is tantamount to understanding the structure of the
Riemann surface associated with the function $w(\xi )$ of the \textit{complex%
} variable $\xi $ defined by the \textit{nondifferential} equation (\ref%
{EqWtilde}), whose different sheets correspond of course to the
different roots of this \textit{nondifferential} equation. In the
\textit{rational} case (\ref{mupq}) this equation is in fact
\textit{polynomial} of degree $J$ in the dependent variable $w$.
Specifically, what must be ascertained is the number of sheets of
this Riemann surface that are accessed by $w(\xi )$ when the
independent variable $\xi $ travels in the \textit{complex} $\xi
$-plane round and round on the circle $\Xi $, whose center
$\bar{\xi}$ and radius $\left\vert R\right\vert $ depend on the
initial data of our physical problem, see (\ref{KSI}) and (18). To
this end one must gain and use a detailed understanding of the
structure of this
Riemann surface. Again, we refer for the details of this analysis to \cite%
{CGSS}. Suffice here to report the following basic information on the
\textit{analytic} structure of the function $w\left( \xi \right) ,$
referring to the general case with \textit{real }$\mu $ (\textit{rational }%
or \textit{irrational}).

The nondifferential equation (\ref{EqWtilde}) defines a $J$-sheeted covering
of the \textit{complex} $\xi $-plane of genus \textit{zero} (of course $J=\infty$ is $\mu$ is irrational). The function $%
w(\xi )$ defined implicitly by this equation features
\textit{square-root} branch points $\xi _{b}$ located on a circle
$B$ centered at the origin of the \textit{complex }$\xi $-plane:
\numparts
\begin{equation}
\xi _{b}=\xi _{b}^{(k)}=r_{b}\,\exp \left( 2\,\pi \,\rmi\,\mu
\,k\right) ~,~~~k=1,2,3,...~,  \label{Singula}
\end{equation}%
\begin{equation}
\xi _{b}=\xi _{b}^{(k)}=r_{b}\,\exp \left[ \rmi\,\frac{2\,\pi
\,p\,k}{q}\right] ~,~\,~k=1,2,...,q~,  \label{Singulb}
\end{equation}%
\begin{equation}
r_{b}=\left( \mu -1\right) ^{\,-1}\,\left( \frac{\mu -1}{\mu }\right)
^{\,\mu }~.  \label{Singulc}
\end{equation}%
\endnumparts
In the last, (\ref{Singulc}), of these formulas it is understood for
definiteness that the \textit{principal} determination is taken of the $\mu $%
-th power appearing in the right-hand side. The first of these formulas, (%
\ref{Singula}), shows clearly that the number of these branch points is
\textit{infinite} if the parameter $\mu $ is \textit{irrational}, and that
they then sit densely on the circle $B$ in the complex $\xi $-plane centered
at the origin and having radius $\left\vert r_{b}\right\vert $, see (\ref%
{Singulc}). Note that this entails that the \textit{generic} point on the
circle $B$ is \textit{not} a branch point (just as a \textit{generic}
\textit{real} number is \textit{not rational}); but every generic point on
the circle $B$ has, if $\mu $ is \textit{irrational}, some branch point (in
fact, an \textit{infinity} of branch points!) \textit{arbitrarily} close to
it (just as every generic \textit{real} number has an \textit{infinity} of
\textit{rational} numbers \textit{arbitrarily} close to it). It is also
important to realize that these branch points are generally on \textit{%
different} sheets of the Riemann surface associated with the function $%
w\left( \xi \right) $: hence, they are dense if one considers the circle $B$
in the \textit{complex} $\xi $-plane, but they are not dense if one
considers these branch points on the Riemann surface itself. As for the
second of these formulas, (\ref{Singulb}), it is instead appropriate to the
case in which the parameter $\mu $ is \textit{rational}, see (\ref{mupq}),
in which case the branch points sit again on the circle $B$ in the complex $%
\xi $-plane, but there are only a \textit{finite} number, $q$, of them (and
note that the factor $p$ appearing in the argument of the exponential in the
right-hand side of this formula, (\ref{Singulb}), is only relevant to
characterize how the branch points $\xi _{b}^{(k)}$ are labeled via the
index $k$). Both in the \textit{irrational} and in the \textit{rational}
case at \textit{all} these branch points the \textit{nondifferential}
equation (\ref{EqWtilde}) has a \textit{double} root which takes the \textit{%
same} value $w(\xi _{b})=\mu $. Note that this entails that circling around
such a branch point in the complex $\xi $-plane corresponds to permuting $2$
of the $J$ roots $w_{j}(\xi )$ among themselves.

In addition, the function $w(\xi )$ possesses branch points at $\xi =0$ and
at $\xi =\infty $, the order of which depends on the value of $\mu $, and is
\textit{rational} if the number $\mu $ is \textit{rational} (see below).

The branch point at $\xi =\infty $ has, if $\mu >1,$ exponent
$-\frac{1}{\mu }$ ($=-\frac{q}{p}$ in the \textit{rational} case),
\begin{equation}
\fl\qquad\quad\xi \approx \infty ~,~\,\,~w(\xi )\approx a\,\xi
^{-1/\mu }\approx 0~,~~~a^{\,\mu }=-\exp (-\rmi\,\pi \,\mu
)~,~~~\mu
>0~, \label{BranchInfa}
\end{equation}%
while it has instead two different exponents if $0<\mu <1$:
\begin{enumerate}
\item the exponent $-%
\frac{1}{\mu }$ ($=-\frac{q}{p}$ in the \textit{rational} case) as given by
the preceding formula (entailing $w\approx 0)$
\item  the exponent $-\frac{1}{%
1-\mu }$ ($=\frac{q}{q-p}$ in the \textit{rational} case) as given by the
following formula (entailing $w\approx 1$),%
\begin{equation}
\fl\qquad\xi \approx \infty ~,~\,\,~w(\xi )\approx 1+a\,\xi
^{-1/(1-\mu )}\approx 1~,~~~a^{\,\mu -1}=1~,~\,~0<\mu <1~.
\label{BranchInfb}
\end{equation}
\end{enumerate}

The branch point at $\xi =0$ is, if $\mu >1,$ of exponent
$\frac{1}{\mu -1}$ ($=\frac{q}{p-q}$ in the \textit{rational}
case),
\begin{equation}
\fl\qquad\quad\xi \approx 0~,~\,\,~w(\xi )\approx 1+a\,\xi
^{1/(\mu -1)}\approx 1~,~~~a^{\,\mu -1}=1~,~~~\mu >1~,
\label{BranchZero}
\end{equation}%
(note the formal analogy of this formula with the previous one, (\ref%
{BranchInfb})), and it is instead \textit{absent} if $0<\mu <1,$ so that in
this second case the \textit{only} branch points in the finite part of the
\textit{complex} $\xi $-plane are those of \textit{square-root} type, see (%
\ref{Singula}). This is the main cause of the difference between
the results,
see below, for this case ($0<\mu <1$) from those for the other two cases ($%
\mu >1$ and $\mu <0)$, which are on the other hand essentially equivalent
among each other being related by the transformation $\mu \mapsto 1-\mu ,$ $%
w(\xi )\mapsto 1-w\left( -\xi \right) ,$ see (\ref{EqWtilde}) (so that
without loss of generality we often forsake an explicit discussion of the
case with $\mu <0$).

Note that these results entail that, in the \textit{rational} case, see (\ref%
{mupq}), making a circle around the branch point at $\xi =0$, in the $p>q$
(i. e. $\mu >1$) case when this branch point is present, causes a cyclic
permutation of $p-q$ of the $p$ roots $w_{j}$: this is particularly evident
if one imagines to travel full circle around the branch point at $\xi =0$ in
its immediate vicinity, since for $\xi \approx 0$ the $p$ roots $w_{j}$ of (%
\ref{EqWtilde}) are clearly divided into \textit{two} sets, a first set of $%
p-q$ roots, disposed equispaced on a circle of small radius ($\approx
\left\vert \xi \right\vert ^{q/(p-q)})$ centered at $w=1$ in the \textit{%
complex} $w$-plane, which then undergo a cyclic permutation among
themselves, and a second set of $q$ roots, disposed equispaced on a circle
in the \textit{complex} $w$-plane centered at the origin and having a large
radius ($\approx \left\vert \xi \right\vert ^{-1}$), each of which after the
operation returns instead to its original position.

The permutation experienced by the $J$ roots $\tilde{w}_{j}(t)$ due to a
sequence of pairwise exchanges of roots -- which take place whenever \textit{%
square-root} branch points are included \textit{inside} the circle $\Xi $
traveled by the point $\xi (t)$ -- causes a reshuffling of the roots which
is nontrivial inasmuch as it depends on how many and on which pairs of roots
get sequentially exchanged over each period, as determined by the number and
identity of \textit{square-root} branch points enclosed inside the circle $%
\Xi $ traveled by $\xi (t)$ and by the detailed structure of the Riemann
surface associated with these branch points, in particular on which sheets
of this Riemann surface the relevant branch points are located. The
reshuffling encompasses more roots when a second mechanism is simultaneously
at play, i. e. that producing a cyclic permutation of $p-q$ roots
(especially, of course, when $p-q$ is large) over each period, as caused by
the presence of the branch point at $\xi =0$: this second mechanism exists
only if $\mu >1$ (or $\mu <0,$ entailing the exchange $-p\mapsto p-q$), and
provided the circle $\Xi $ traveled by $\xi (t)$ \textit{does include} the
point $\xi =0.$ This phenomenology causes the possible periods $\tilde{T}=%
\tilde{j}\,T$ of the time evolution of the \textit{generic }root $\tilde{w}%
(t)$ to depend on the initial data, but remarkably we will see that, for
given initial data, there are (at most) only $3$ possible values of these
periods, and that they can be given explicitly in terms of the initial data
and of the two numbers $p$ and $q,$ see (\ref{mupq}). Indeed we show below
that analogous results can as well be given in the case when $\mu $ is
\textit{irrational}, in spite of the fact that the dependence on the initial
data may then be quite \textit{sensitive}.

The correspondence of the analysis, given here in terms of the dynamics of
the roots $\tilde{w}_{j}(t)$, with that in terms of travel on the Riemann
surface made above (see paragraphs after (\ref{CondIso})), including the
distinction made there about \textit{active }and \textit{inactive} branch
points, should be noted: the \textit{active }branch points are those that
cause a reshuffling of roots that involves the ``physical'' root $\check{w}(t)$%
, the \textit{inactive} ones are those that do \textit{not} cause
a reshuffling of roots that involves the ``physical'' root
$\check{w}(t)$,
either because they cause no reshuffling at all being located \textit{%
outside }the relevant circle ($\Xi $ in the \textit{complex} $\xi $-plane in
the context of the present analysis, $C$ in the \textit{complex} $\tau $%
-plane in the context of the discussion made above when the distinction
among \textit{active} and \textit{inactive} branch points was first
introduced), or because they cause a reshuffling which however does not
involve\textit{\ }the physical root $\check{w}(t)$.

We now report our findings concerning the time evolution of a generic root $%
\tilde{w}(t)$, referring at first mainly to the \textit{rational}
case but including immediately results for the \textit{irrational}
case whenever it is convenient to do so in order to shorten our
presentation. The remaining information on the \textit{irrational}
case is provided below (see Proposition 5).

First of all it is useful to visualize the two circles $B$ and $\Xi $ in the
\textit{complex }$\xi $-plane (draw them!): recall that the circle $B$ on
which the branch points sit is centered at the origin and its radius $%
\left\vert r_{b}\right\vert $ only depends on the parameter $\mu ,$ see (\ref%
{Singulc}), while both the center $\bar{\xi}$ and the radius
$\left\vert
R\right\vert $ of the circle $\Xi $ traveled upon by $\xi (t),$ see (\ref%
{KSI}), \textit{do depend} on the initial data, see (18).

\textbf{Proposition 1. }
If the circle $\Xi $ is \textit{inside} the circle $%
B $ (i. e. $\left\vert \bar{\xi}\right\vert +\left\vert
R\right\vert <\left\vert r_{b}\right\vert ,$ see (18) and
(\ref{Singulc})), and (a) $\mu $ is \textit{inside} the interval
$0<\mu <1$ or (b) $\mu $ is \textit{outside} this interval ($\mu
>1$ or $\mu <0)$ and moreover the
circle $\Xi $ does \textit{not} include the origin $\xi =0$ (i. e. $%
\left\vert R\right\vert <\left\vert \bar{\xi}\right\vert $ or equivalently $%
\left\vert \eta \right\vert >1$), then $\tilde{j}=1,$ i. e. the \textit{%
generic }solution $\tilde{w}(t)$ is periodic with period $T,$ $\tilde{w}%
(t+T)=\tilde{w}(t)$. This outcome applies equally if $\mu $ is \textit{%
rational} or \textit{irrational}. $\square $

\textbf{Remark 1}. In \textit{all} the cases identified in this
Proposition 1 there are \textit{no branch points at all} inside the circle $%
\Xi $: indeed the outcome detailed by this Proposition 1 applies
in \textit{all} the cases in which this happens (see our
discussion above), even (in the case with \textit{rational} $\mu
)$ if the circles $B$ and $\Xi $ do cross each other marginally;
and of course in \textit{all} these cases
\textit{all} the roots $\tilde{w}_{j}(t)$ are \textit{periodic} with period $%
T,$ and in the context of the physical problem (\ref{EqMot}) the solution is
characterized by the simple periodicity rule (\ref{Periodic}). The
restriction (\ref{CondIso}) on the initial data is \textit{sufficient} (but
of course not \textit{necessary}) to guarantee that we are in this regime.$%
\square $

\textbf{Proposition 2}. If the circle $\Xi $ is \textit{inside} the circle $%
B $ (i. e. $\left\vert \bar{\xi}\right\vert +\left\vert
R\right\vert <\left\vert r_{b}\right\vert $ ),  the circle $\Xi $
\textit{does} include the origin $\xi =0$ (i. e. $\left\vert
R\right\vert >\left\vert \bar{\xi}\right\vert $ or equivalently
$\left\vert \eta \right\vert <1$), and $\mu $ is \textit{outside}
the interval $0<\mu <1$ then in the \textit{rational} case, see (\ref{mupq}%
),
\begin{enumerate}
\item[(a)] $\tilde{j}=1$ or $\tilde{j}=p-q$ if $\mu>1$ \item[(b)]
$\tilde{j}=1$ or $\tilde{j}=\left\vert p\right\vert $ if $\mu <0$.
\end{enumerate}
In the \textit{irrational case} the time evolution of the \textit{%
generic} root $\tilde{w}(t)$ is either \textit{periodic }with the basic
period $T,$ or \textit{quasiperiodic}, involving in particular a (nonlinear)
superposition of \textit{two} periodic evolutions with \textit{two}
noncongruent periods, specifically (a) with period $T$ and $\frac{T}{\mu -1}$
if $\mu >1,$ (b) with period $T$ and $\frac{T}{\left\vert \mu \right\vert }$
if $\mu <0$.$\square $

\textbf{Remark 2}. In the case identified in this Proposition 2
the only branch point \textit{inside }$\Xi $ is that at $\xi =0,$
which is indeed only present if $\mu >1$ or $\mu <0.$ Hence in the
\textit{rational} case with $p>q$ (i. e. $\mu >1)$, $p-q$ roots
$\tilde{w}_{j}(t)$ get cyclically exchanged among themselves,
entailing that the time evolution of each of them has period
$\left( p-q\right) \,T,$ while the remaining $q$ roots have period
$T;$ with an analogous phenomenology in the $\mu <0$. Likewise,
when $\mu $ is \textit{irrational,} the periodicity of the time
evolution of the \textit{generic} root $\tilde{w}(t)$ has period
$T$ if the branch point at $\xi =0$ is \textit{inactive} (i. e.,
it does not appear on the sheet on which $\tilde{w}(0)$ lives),
otherwise its time evolution can be inferred by replacing $\xi $
with $\xi (t),$ see (\ref{KSI}), in the formula characterizing the
branch point at $\xi =0$, see (\ref{BranchZero}). Note however
that for the \textit{special} initial data such that the two
circles $B$ and $\Xi $ are \textit{concentric} (i. e., $\eta =0$)
the
\textit{quasiperiodic }time evolution of $\tilde{w}(t)$ is instead \textit{%
periodic} (a) with period $\frac{T}{\mu -1}$ if $\mu >1,$ (b) with period $%
\frac{T}{\left\vert \mu \right\vert }$ if $\mu <0$.$\square $

\textbf{Proposition 3}. If the circle $\Xi $ is  \textit{outside}
the
circle $B$ (i. e. $\left\vert R\right\vert >\left\vert \bar{\xi}%
\right\vert +\left\vert r_{b}\right\vert $ ) then in the \textit{rational} case, see (\ref{mupq}),
\begin{enumerate}
\item[(a)] $\tilde{j}%
=p$ if $\mu>1$, \item[(b)] $\tilde{j}=q-p=q+\left\vert
p\right\vert $ if $\mu<0$ \item[(c)] $\tilde{j}=p$ or
$\tilde{j}=q-p$ if $0<\mu <1$.
\end{enumerate}
 In the \textit{irrational case} the
time
evolution of the \textit{generic} root $\tilde{w}(t)$ is \textit{%
quasiperiodic}, involving a (nonlinear) superposition of \textit{two}
periodic evolutions with \textit{two} noncongruent periods, specifically
\begin{enumerate}
\item[(a)]
with periods $T$ and $\frac{T}{\mu }$ if $\mu >1,$
\item[(b)] with periods $T$ and $%
\frac{T}{1-\mu }$ if $\mu <0,$
\item[(c)] with periods $T$ and $\frac{T}{\mu }$ or $%
T$ and $\frac{T}{1-\mu }$ if $0<\mu <1$.$\square $
\end{enumerate}
\textbf{Remark 3. }In all the cases encompassed in this
Proposition 3  the branch points of $w\left( \xi \right) $ in the
\textit{finite} part of the \textit{complex }$\xi $-plane are
\textit{all inside }the circle $\Xi $, hence the dynamics of the
roots $\tilde{w}_{j}(t)$ can be understood in
terms of the branch point of $w\left( \xi \right) $ at $\xi =\infty $. Therefore in the \textit{rational }case with $p>q$ (i. e. $%
\mu >1$) \textit{all} the $p$ roots get cyclically exchanged, so that each
of them gets back to its original value after a period $p\,T$; likewise if $%
p<0$ (i. e. $\mu <0$) \textit{all} the $q+\left\vert p\right\vert $ roots
get cyclically exchanged, so that each of them gets back to its original
value after a period $\left( q+\left\vert p\right\vert \right) \,T.$ In the
other \textit{rational }case, $0<p<q$ (i. e. $0<\mu <1$), $p$ roots gets
cyclically exchanged among themselves, and the remaining $q-p$ roots get
cyclically exchanged among themselves, so that the \textit{generic} root $%
\tilde{w}(t)$ has period $p\,T$ if it belongs to the first set, and $\left(
q-p\right) \,T$ if it belongs to the second. And the outcome in the \textit{%
irrational} case can as well be understood in terms of the
exponent of the branch point at $\xi =\infty ,$ see
(\ref{BranchInfa})-(\ref{BranchInfb}).$\square $

The situation is less straightforward (hence more interesting) if
the two circles $B$ and $\Xi $ \textit{do intersect} each other
(i. e. $\left\vert \bar{\xi}\right\vert -\left\vert R\right\vert
<\left\vert r_{b}\right\vert <\left\vert \bar{\xi}\right\vert
+\left\vert R\right\vert $). Then the parameter that plays a
crucial role is the number $b$ of \textit{square-root }branch
points, sitting on the circle $B,$ that fall \textit{inside} the
circle $\Xi $. This number $b$ is \textit{finite}, $1\leq b\leq q$
(note that the case $b=0$ is taken care of by Proposition 1) only
in the \textit{rational} case, to which we restrict consideration
in the following Proposition 4 (and we exclude from consideration
the \textit{nongeneric} case in which the circle $\Xi $ hits one
of the branch points sitting on the circle $B$). Then \textit{two}
or \textit{three} (but no more!) different alternatives are
possible for the value of the \textit{positive integer
}$\tilde{j},$ as detailed below.

Hereafter the notation $\left\lfloor x\right\rfloor $ denotes the \textit{%
floor} of the \textit{real} number $x,$ namely the \textit{largest integer
number not larger than} $x$ (hence for instance $\left\lfloor
-0.3\right\rfloor =-1,$ $\left\lfloor 0\right\rfloor =0)$.

\textbf{Proposition 4}. (i). If $p>q$ (i. e. $\mu >1,$ see
(\ref{mupq})) and the origin $\xi =0$ is \textit{outside} the
circle $\Xi $ (i. e. $\left\vert \eta \right\vert >1$), then
$\tilde{j}$ can take one of the following $3$ values:
\begin{equation}
\tilde{j}=1~\,~\mbox{or~~\thinspace \thinspace }\tilde{j}=\left\lfloor \frac{%
b-1}{p-q}\right\rfloor +1~\,~\mbox{or~~\thinspace \thinspace }\tilde{j}%
=\left\lfloor \frac{b-1}{p-q}\right\rfloor +2~.  \label{jtilaa}
\end{equation}

(ii). If $p>q$ (i. e. $\mu >1,$ see (\ref{mupq})) and the origin $\xi =0$ is
\textit{inside} the circle $\Xi $ (i. e. $\left\vert \eta \right\vert <1$),
then $\tilde{j}$ can take one of the following $2$ values:%
\begin{equation}
\tilde{j}=1~\,~\mbox{or~~\thinspace \thinspace }\tilde{j}=b+p-q~.
\label{jtilab}
\end{equation}%
\qquad

(i'). If $p<0$ (i. e. $\mu <0,$ see (\ref{mupq})) and the origin
$\xi =0$ is \textit{outside} the circle $\Xi $ (i. e. $\left\vert
\eta \right\vert >1$), then $\tilde{j}$ can take one of the
following $3$ values:
\begin{equation}
\tilde{j}=1~\,~\mbox{or~~\thinspace \thinspace }\tilde{j}=\left\lfloor \frac{%
b-1}{\left\vert p\right\vert }\right\rfloor
+1~\,~\mbox{or~~\thinspace
\thinspace }\tilde{j}=\left\lfloor \frac{b-1}{\left\vert p\right\vert }%
\right\rfloor +2~.  \label{jtilba}
\end{equation}

(ii'). If $p<0$ (i. e. $\mu <0,$ see (\ref{mupq})) and the origin $\xi =0$
is \textit{inside} the circle $\Xi $ (i. e. $\left\vert \eta \right\vert <1$%
), then $\tilde{j}$ can take one of the following $2$ values:%
\begin{equation}
\tilde{j}=1~\,~\mbox{or~~\thinspace \thinspace
}\tilde{j}=b+\left\vert p\right\vert ~.  \label{jtilbb}
\end{equation}

(iii). The situation is more intriguing if $0<p<q$ (i. e., $0<\mu <1,$ see (%
\ref{mupq})). Then one must introduce the \textit{simple continued
fraction} expansion of the number
\begin{equation}
\frac{q}{q-p}=\frac{1}{1-\mu }=a_{0}+\frac{1}{a_{1}+\frac{1}{a_{2}+...}}~.
\label{ConFuna}
\end{equation}%
The $k$th \textit{convergent} $C_{k}$ of this \textit{continued fraction }%
expansion (\ref{ConFuna}), and in particular its numerator $P_{k}$ and
denominator $Q_{k}$,%
\begin{equation}
C_{k}=\frac{P_{k}}{Q_{k}}~,  \label{ConFund}
\end{equation}%
are then defined by the recursions%
\begin{eqnarray}
\fl\qquad P_{k}=a_{k}\,P_{k-1}+P_{k-2}~,~~~ &P_{-2}=0~,~~~ P_{-1}=1~,~~~\quad k=0,1,2,\dots\\
\label{ConFune} \fl\qquad Q_{k}=a_{k}\,Q_{k-1}+Q_{k-2}~,~~~
&Q_{-2}=1~,~~~ Q_{-1}=0,~~~\quad k=0,1,2,\dots \label{ConFunf}
\end{eqnarray}%
Note that these formulas apply equally if $\mu $ is \textit{%
rational} or \textit{irrational}; of course depending whether $\mu $ is
\textit{rational} or \textit{irrational} the \textit{continued fraction}
expansion does or does not terminate. In the \textit{rational} case under
present discussion we also introduce another sequence of \textit{nonnegative
integers}:

\begin{equation}
b_{k}=q-(-)^{k}\,\left[ \left( p-q\right)
\,P_{k-2}+q\,Q_{k-2}\right] ~,~~~ \quad k=0,1,2,\dots  \label{bka}
\end{equation}%
Given $b$, let the integer $h$ and the period $T(b)$ be defined by
the following formulas:
\begin{equation}
 b_{h}\le b<b_{h+1},
 \end{equation}
\begin{equation}
\label{T2}
\begin{array}{l}
T(b)=
P_{h-2}+\left(\left\lfloor\frac{b-b_h-1}{q-b_{h+1}}\right\rfloor
+2 \right)P_{h-1}.
\end{array}
\end{equation}
Then the roots of the polynomial can have only one of the
following three periods $\tilde j$:
\begin{equation}
\tilde j = T(b)\quad\mbox{or}\quad \tilde j =
T(b)-P_{h-1}\quad\mbox{or}\quad \tilde j = P_{h-1}. \label{jtilc}
\end{equation}
 This is the
generic case; there are however some cases in which the roots have
only the two periods $T(b)$ and $P_{h-1}$. This happens of course
when $T(b)=2 P_{h-1}$ and whenever $b$ takes the following special
values: \begin{equation}b=b_h+n(q-b_{h+1}),\qquad 0\le n\le
a_h-1,\quad n\,\,\mbox{integer}.\qquad \square\end{equation}

\textbf{Remark 4. }In case (i) of Proposition 4 the mechanism that
yields periods longer than unity is the coming into play of the $b$ \textit{%
square-root} branch points enclosed \textit{inside} the circle $\Xi $, which
cause a certain number of roots $\tilde{w}_{j}(t)$ to exchange pairwise
their roles through the time evolution. But this phenomenology only affects
some roots; others remain unaffected, hence their periods remain unity, and
this explains the first entry in (\ref{jtilaa}). The precise form of the
other entries in this formula, (\ref{jtilaa}), requires of course a more
detailed treatment, see \cite{CGSS}; the outcome there depends on how many
pair exchanges actually do take place, or, equivalently, how many sheets of
the Riemann surface get actually visited, and this depends in a fairly
detailed manner on the specific structure of this surface. But note that
only \textit{two} different periods may emerge, differing by only \textit{one%
} unit.

In case (ii) of Proposition 4 the second mechanism, associated
with the presence of the branch point at $\xi =0,$ comes
additionally into play, causing the connection of \textit{all} the
$b$ sheets containing the $b$
\textit{square-root }branch points, both among themselves and with the $%
(p-q) $ sheets containing the branch point at $\xi =0$. The corresponding $%
\left( b+p-q\right) $ roots get permuted among themselves through the time
evolution, with the period indicated by the second entry in (\ref{jtilab})
(note that whenever $p$ is \textit{quite large} and $b$ is \textit{close} to
its \textit{maximal} value $q-1,$ the resulting period is \textit{quite
large)}. The remaining $(q-b)$ sheets, corresponding to the $\left(
q-b\right) $ branch points lying on the circle $B$ \textit{outside} the
circle $\Xi $, are isolated, hence the corresponding roots do \textit{not}
take part in the quadrille, so that their period remains unity, as indicated
by the first entry in (\ref{jtilab}).

The cases (i') and (ii') of Proposition 4 require no additional
discussion.

In case (iii) of Proposition 4 (with $0<p<q$, i. e. $0<\mu <1$)
there is no branch point at $\xi =0$; hence the mechanism is now \textit{%
absent }that previously caused the connection of all the sheets associated
with the $b$ \textit{square-root} branch points sitting on the circle $B$
\textit{inside} the circle $\Xi $. This implies, see \cite{CGSS}, that each
sheet of the Riemann surface contains \textit{only two} branch points, whose
projections on the circle $B$ are separated by $(p-q)$ other branch points.
Note that this entails that two branch points which are \textit{adjacent} on
the circle $B$ in the \textit{complex} $\xi $-plane are instead
topologically far apart on the Riemann surface of the function $w(\xi ),$
living on sheets which are not directly connected. As a consequence the
period of the time evolution of the root $\tilde{w}(t)$ does not change, as
the initial data change causing the circle $\Xi $ to change its position and
dimension so that the number $b$ of \textit{square-root} branch points
enclosed in it increases, until some crucial \textit{square-root }branch
point gets thereby included inside the circle $\Xi ,$ causing the connection
of two separate groups of connected sheets; and this mechanism occurs more
and more frequently as $b$ increases more and more. This explained \textit{%
qualitatively} the piecewise constant behavior of the period as
$b$ increases, characterized by shorter and shorter steps and by
bigger and bigger jumps, see (\ref{T2})-(\ref{jtilc}). The exact
treatment of this mechanism, yielding (\ref{T2})-(\ref{jtilc}), is
rather complicated, as indicated by the role
played by the \textit{continued fraction }expansion of the number $\frac{q}{%
q-p}$. The details are given in \cite{CGSS}. Here we
limit ourselves to emphasizing that, for given initial data, the \textit{%
generic} root $\tilde{w}(t)$ can have only $3$ possible periods, the third
of which is just the sum of the first two, see (\ref{jtilc}).$\square $

Let us now discuss the case in which $\mu $ is an \textit{irrational}
number, recalling that we are now considering initial data ($\left\vert \bar{%
\xi}\right\vert -\left\vert R\right\vert <\left\vert
r_{b}\right\vert <\left\vert \bar{\xi}\right\vert +\left\vert
R\right\vert $) such that the two circles $B$ and $\Xi $
\textit{do intersect} each other (the results for the other cases
have been given in Propositions $1$--$3$. ). One must then
introduce the ratio $\nu $ of the length of the arc of the circle
$B$ that is \textit{inside} the circle $\Xi ,$ to the length of
the \textit{entire} circle $B$: of course $0<\nu <1.$ Expressing
this ratio in terms of the initial data is an exercise in
elementary plane geometry
yielding the formula%
\begin{equation}
\nu =\frac{1}{\pi }\,\arccos \left( \frac{\left\vert r_{b}\right\vert
^{\,2}+\left\vert \bar{\xi}\right\vert ^{\,2}-\left\vert R\right\vert ^{\,2}%
}{2\,\left\vert r_{b}\,\bar{\xi}\right\vert }\right) ~,  \label{nu}
\end{equation}%
where the determination of the $\arccos $ function must be chosen so that $%
0<\nu <1.$ The results for the periods are then given by the
following

\textbf{Proposition 5}.
\noindent (i) If $\mu >1$ and the origin $\xi =0$ is \textit{%
outside} the circle $\Xi $ (i. e. $\left\vert \eta \right\vert >1$), the
time evolution of the \textit{generic} root $\tilde{w}(t)$ is still \textit{%
periodic} with period $\tilde{T}=$ $\tilde{j}\,T$ and $\tilde{j}$
can take one of the following $3$ values:
\begin{equation}
\tilde{j}=1~\,~\mbox{or~~\thinspace \thinspace }\tilde{j}=\left\lfloor \frac{%
\nu }{\mu -1}\right\rfloor +1~\,~\mbox{or~~\thinspace \thinspace }\tilde{j}%
=\left\lfloor \frac{\nu }{\mu -1}\right\rfloor +2~.  \label{Jtilaa}
\end{equation}%
Note that the second and third entry only differ by one unit and moreover
that, if $\mu >2,$ the floor functions vanish, hence for \textit{all} values
of $\mu $ larger than $2$ (and $\left\vert \eta \right\vert >1$) the only
possible values for $\tilde{j}$ are $1$ or $2.$

(ii) If $\mu >1$ and the origin $\xi =0$ is \textit{inside} the
circle $\Xi $ (i. e. $\left\vert \eta \right\vert <1$), the time
evolution of the \textit{generic} root $\tilde{w}(t)$ is either
\textit{periodic} with period $T$ or \textit{aperiodic}.

(i'). If $\mu <0$ and the origin $\xi =0$ is \textit{outside} the circle $%
\Xi $ (i. e. $\left\vert \eta \right\vert >1$), the time evolution of the
\textit{generic} root $\tilde{w}(t)$ is still \textit{periodic} with period $%
\tilde{T}=$ $\tilde{j}\,T$ and $\tilde{j}$ can take one of the
following $3$ values:
\begin{equation}
\tilde{j}=1~\,~\mbox{or~~\thinspace \thinspace }\tilde{j}=\left\lfloor \frac{%
\nu }{\left\vert \mu \right\vert }\right\rfloor
+1~\,~\mbox{or~~\thinspace
\thinspace }\tilde{j}=\left\lfloor \frac{\nu }{\left\vert \mu \right\vert }%
\right\rfloor +2~.  \label{Jtilba}
\end{equation}%
Again the second and third entry only differ by one unit and moreover if $%
\left\vert \mu \right\vert >2$ the floor functions vanish, hence we conclude
that for \textit{all} values of $\mu $ smaller than $-2$ (and $\left\vert
\eta \right\vert >1$) the only possible values for $\tilde{j}$ are $1$ or $%
2. $

(ii') If $\mu <0$ and the origin $\xi =0$ is \textit{inside} the circle $\Xi
$ (i. e. $\left\vert \eta \right\vert <1$), the time evolution of the
\textit{generic} root $\tilde{w}(t)$ is either \textit{periodic} with period
$T$ or \textit{aperiodic}.

(iii). The case $0<\mu <1$ is again more intriguing, and it
requires again the use of the \textit{continued fraction}
expansion of $\frac{1}{1-\mu }$, which however now does
\textit{not} terminate. We define now in addition the (endless)
sequence of \textit{real} numbers
\begin{equation}
\nu _{k}=1-(-)^{\,k}\,\left[ \left( \mu -1\right) \,P_{k-2}+Q_{k-2}\right]
,~~~k=0,1,2,...  \label{nuka}
\end{equation}%
(entailing $\nu _{0}=0$), and we then identify the \textit{nonnegative
integer} $h$ via the inequalities%
\begin{equation}
0\leq \nu _{h}\leq \nu <\nu _{h+1}<1~.  \label{nukb}
\end{equation}%
Let $T(\nu)$ be defined by the following expression:
\begin{equation} T(\nu)=
P_{h-2}+\left(\left\lfloor\frac{\nu-\nu_h}{1-\nu_{h+1}}\right\rfloor
+2 \right)P_{h-1}.\end{equation}
Then the motion of the \textit{generic }root $\tilde{w}(t)$ is again \textit{%
periodic} with period $\tilde{T}=\tilde{j}\,T$ where $\tilde{j}$
can take one of the following $3$ values:
\begin{equation}
\tilde j = T(\nu)\quad\mbox{or}\quad \tilde j =
T(\nu)-P_{h-1}\quad\mbox{or}\quad \tilde j = P_{h-1}.\qquad\quad
\square \label{Jtilc}
\end{equation}
\textbf{Remark 5}. The results for this case with
\textit{irrational} $\mu $ can be obtained from those for
\textit{rational }$\mu $ (see Proposition 4) by taking
appropriately the limit in which
\begin{enumerate}
\item[(a)] the integers $p$ and $q$ diverge with their ratio $\mu
$ fixed (see (\ref{mupq})), and \item[(b)] the number $b$ of
\textit{square-root} branch points \textit{inside} the
circle $\Xi ,$ as well as the \textit{total} number $q$ of \textit{%
square-root} branch points, also diverge with their ratio fixed (recall that
\textit{all} these \textit{square-root} branch points sit, densely
equispaced, on the circle $B$ in the \textit{complex }$\xi $-plane although
on \textit{different} sheets of the Riemann surface of the function $w\left(
\xi \right) $ -- hence this ratio coincides with the quantity $\nu $
defined, and evaluated in terms of the initial data, above, see (\ref{nu})).
\end{enumerate}
This also suggests obvious extensions to the present case with \textit{%
irrational }$\mu $ of comments contained in the Remark 4, which
will not be repeated here. We therefore limit below our remarks to
aspects of the results reported in Proposition 5 having no
immediate counterpart in the comments contained in Remark 4.

In the cases (i) respectively (i') of Proposition 5 the rules
giving the period of the time evolution of the \textit{generic} root $\tilde{%
w}(t),$ see (\ref{Jtilaa}) respectively (\ref{Jtilba}), are fairly
straightforward and generally yield rather small periods, unless $\mu
=1+\varepsilon $ respectively $\mu =-\varepsilon $ with $\varepsilon $ an
\textit{irrational} number \textit{positive }but \textit{extremely small}.

In cases (ii) and (ii') of Proposition 5 the situation is quite
interesting because the time evolution of the generic root
$\tilde{w}(t)$
can be either \textit{periodic} with the basic period $T$ or \textit{%
aperiodic}. Note that in these cases the circle $\Xi $ intersects the circle
$B$ that is \textit{densely} filled with \textit{square-root} branch points,
and moreover the branch point at $\xi =0$ (which is now of \textit{%
irrational }exponent, see (\ref{BranchZero})) is \textit{inside} the circle $%
\Xi .$ This entails that, of the \textit{infinity} of \textit{square-root}
branch points located on the piece of the circle $B$ that is \textit{inside}
$\Xi ,$ either \textit{none}, or \textit{all}, are \textit{active}. The
first case obtains if the root $\tilde{w}(t)$ under consideration is \textit{%
initially} on a sheet containing a branch point that does \textit{not} fall
inside $\Xi ,$ hence the time evolution of this root $\tilde{w}(t)\equiv w%
\left[ \xi (t)\right] $ as the point $\xi (t)$ travels round and round on
the circle $\Xi $ brings it back to its point of departure after a single
round; equivalently, in this case the root $\tilde{w}(t)\equiv w\left[ \xi
(t)\right] $ belongs to a set of roots that does \textit{not} get permuted
as the point $\xi (t)$ travels round and round on the circle $\Xi $. In the
second case the root $\tilde{w}(t)$ under consideration starts from a sheet
of the Riemann surface that contains a branch point inside $B,$ so that,
when $\xi (t)$ travels round and round on the circle $\Xi $, an \textit{%
endless} sequence of \textit{different} sheets get accessed by $\tilde{w}%
(t)\equiv w\left[ \xi (t)\right] $; equivalently, such a root $\tilde{w}%
(t)\equiv w\left[ \xi (t)\right] $ belongs to a set (including an \textit{%
infinity }of roots) that does get permuted as the point $\xi (t)$ travels
round and round on the circle $\Xi $, with both mechanisms -- the pairwise
exchange of some roots, and the cyclic permutation of an \textit{infinite }%
number of roots -- playing a role at each round. The identification of which
sheets get thereby accessed, and in which order -- namely the specific shape
of the trajectory when looked at, as it were stroboscopically, at the
discrete sequence of instants $T_{k}=k\,T,$ $k=1,2,3,...$-- is discussed in
\cite{CGSS}. The extent to which this regime yields \textit{irregular }%
motions is discussed further below, also to illuminate the distinction in
these regimes between the time evolution entailed by our model with a given
\textit{irrational} value of $\mu ,$ and that of the analogous models with
\textit{rational} values of $\mu $ providing more and more accurate
approximations of the given \textit{irrational }value of $\mu $.

In case (iii) the time evolution is still \textit{isochronous}, inasmuch as
the results reported above entail that, for \textit{any} given initial data
(excluding, of course, the \textit{special} ones leading to a collision;
which are \textit{special} in the same sense as a \textit{rational} number
is \textit{special} in the context of \textit{real} numbers), the motion of
every root $\tilde{w}_{j}(t)$ is \textit{periodic} with one of the $3$
periods entailed by (\ref{Jtilc}) (and note that the value of the integer $%
\tilde{j}$ provided by the \textit{third} of these $3$ formulas is just the
sum of the $2$ values for $\tilde{j}$ provided by the first $2$ of these $3$
formulas). It is indeed clear that the initial data yielding such an outcome
are included in an \textit{open} set of such data, having of course \textit{%
full dimensionality} in the space of initial data, \textit{all} yielding the
\textit{same} outcome: since the periods do \textit{not} change, see (\ref%
{Jtilc}), if the change of the initial data, hence the change in the ratio $%
\nu ,$ is sufficiently tiny. However the measures of these sets of data
yielding the \textit{same} outcome gets progressively \textit{smaller} as
the predicted periods get \textit{larger}, and moreover the corresponding
predictions involve more and more terms in the (never ending) \textit{%
continued fraction }expansion of the irrational number
$\frac{1}{1-\mu }$, see (\ref{ConFuna}), displaying thereby, as
$\nu $ increases towards unity, a \textit{progressively more
sensitive }dependence of the periodicity of our system on the
initial data and moreover on the parameters (the coupling
constants, that determine the value of $\mu $, see (\ref{mu})) of
our physical model (\ref{EqMot}).$\square $

\vskip 0.3cm\noindent\textbf{Example}. Let us display here a
specific example with the following
(conveniently chosen) \textit{irrational} value of $\mu $ in the interval $%
0<\mu <1$ (hence corresponding to case (iii) of Proposition 5):
\begin{equation}
\mu =\frac{2}{3+\sqrt{5}}=\frac{1}{1+\varphi }~,~~~\frac{1}{1-\mu }=\varphi =%
\frac{1+\sqrt{5}}{2}~,  \label{Aureaa}
\end{equation}%
where of course $\varphi $ is the \textit{golden ratio}. This assignment
entails that \textit{all} the coefficients $a_{k}$, see (\ref{ConFuna}), are
in this case unity, $a_{k}=1$, hence the quantities $P_{k},$ see (\ref%
{ConFune}), coincide with the Fibonacci numbers $F_{k}$,%
\begin{eqnarray}
&P_{k}=F_{k}~,~~\,F_{-2}=0~,~~~F_{-1}=1~,~~~F_{0}=1~,\\
& F_{k+2}=F_{k}+F_{k+1}~,  \label{Aureab}
\end{eqnarray}%
and moreover one easily finds that
\begin{equation}
\nu _{k}=1-\varphi ^{\,-k}~.  \label{Aureac}
\end{equation}%
The corresponding formula for the periods obtains inserting these values in (%
\ref{nukb}) and (\ref{Jtilc}). From it, with some labor, one can obtain the
following controlled estimate for the possible values of $\tilde{j}$:%
\begin{equation}
\frac{\nu \,\left( 2-\nu \right) }{\sqrt{5}\,\left( 1-\nu \right) }\leq
\tilde{j}\leq \frac{\sqrt{5}+2-\left( \sqrt{5}+1\right) \,\nu \,(2-\nu )}{%
\left( \sqrt{5}-1\right) \,\left( 1-\nu \right) }~.  \label{Auread}
\end{equation}%
These inequalities are valid for all values of $\nu $ (in the interval $%
0<\nu <1$); they clearly entail that the integer $\tilde{j}$
diverges proportionally to $\left( 1-\nu \right) ^{-1}$ as $\nu
\rightarrow 1,$ and that for $0<\nu <\bar{\nu}$ with
\begin{equation}
\bar{\nu}=\sqrt{5}-2+\frac{3-\sqrt{5}}{2}\,\sqrt{3-\frac{\sqrt{5}}{2}}%
=0.760067\dots  \label{Aureae}
\end{equation}%
the only possible values of $\tilde{j}$ are $2$ and $3.\square $

This concludes our presentation of the results, and of some related
observations, detailing the periodicity (if any) of the time evolution of a
\textit{generic} root $\tilde{w}(t)$ of (\ref{EqWtilde}) with (\ref{KSI}).
The identification of analogous, but of course more definite, results for
the \textit{physical} root $\check{w}(t),$ and the consequential information
on the periodicity (if any) of the solution of the physical problem (\ref%
{EqMot}) -- as well as some additional information on the corresponding
trajectories of the coordinates $z_{n}(t)$ -- are provided in \cite{CGSS}.

Last but not least let us elaborate on the character of the \textit{aperiodic%
} time evolution indicated under item (ii) of Proposition 5
, including the extent it is \textit{irregular} and it depends \textit{%
sensitively} on its initial data. It is illuminating to relate
this question with the finding reported under item (ii) of
Proposition 4, also in order to provide a better understanding of
the relationship among
the \textit{aperiodic} time evolution that can emerge when $\mu $ is \textit{%
irrational} (see item (ii) of Proposition 5) and the corresponding
behavior -- say, with the same initial data -- for a sequence of
models with \textit{rational} values of $\mu $ (see (\ref{mupq}))
that provide better and better approximations to that
\textit{irrational }value of $\mu $; keeping in mind the
\textit{qualitative} difference among the
\textit{aperiodic} time evolution emerging when $\mu $ is \textit{irrational}%
, and the \textit{periodic} -- indeed, even \textit{isochronous}
-- time evolutions prevailing whenever $\mu $ is
\textit{rational}, albeit with the qualifications indicated under
item (ii) of Remark 4. Note that we are now discussing the case
$\mu >1$ (an analogous discussion in the $\mu <0$
case can be forsaken), with initial data such that the two circles $B$ and $%
\Xi $ in the \textit{complex }$\xi $-plane \textit{do intersect} and
moreover the origin $\xi =0$ falls \textit{inside} the circle $\Xi $ (i. e. $%
\left\vert r_{b}\right\vert <\left\vert \bar{\xi}\right\vert +\left\vert
R\right\vert $ and $\left\vert \eta \right\vert <1).$

Let us then consider a given \textit{irrational }value of $\mu >1$ and let
the \textit{rational} number $\frac{p}{q}$ (with $p>q$) provide a \textit{%
very good }approximation to $\mu ,$ which of course entails that the \textit{%
positive integers} $p$ and $q$ are both \textit{very large}.
Consider then the \textit{difference}
\begin{equation}
\Delta \,\tilde{j}=\Delta \,b  \label{DeltaTa}
\end{equation}%
(see the second entry in (\ref{jtilab})) between the \textit{two} positive
integers that characterize the \textit{two} periods of the \textit{two }time
evolutions of $\tilde{w}(t)$ corresponding to \textit{two} sets of initial
data that differ \textit{very little}. Here clearly the quantity $\Delta \,b$
is the \textit{difference} between the number of branch points that are
enclosed inside the circle $\Xi $ for these \textit{two} different sets of
initial data. Since the number $q$ of branch points on $B$\thinspace is
\textit{very large}, it stands to reason that%
\begin{equation}
\Delta \,b=\left\lfloor O(q\,\delta )\right\rfloor ~,  \label{DeltaTb}
\end{equation}%
where the \textit{quite small} (\textit{positive})\textit{\ }number $\delta $
provides a (dimensionless) measure of the difference between the \textit{two}
sets of initial data (see for instance (\ref{nu}), which clearly becomes
approximately applicable when $q$ is \textit{very large}). The \textit{floor}
symbol $\left\lfloor \,\right\rfloor $ has been introduced in the right-hand
side of this formula to account for the integer character of the numbers $b$
hence of their difference $\Delta \,b,$ while the \textit{order of magnitude}
symbol $O\left( \,\right) $ indicates that the difference $\Delta \,b$ is
proportional (in fact equal, given the latitude left by our definition of
the quantity $\delta )$ to the quantity $q\,\delta $ up to \textit{%
corrections} which become \textit{negligibly small} when $\delta $ is
\textit{very small} and $q$ is \textit{very large}, but irrespective of the
value of the quantity $q\,\delta $ itself which, as the product of the
\textit{large} number $q$ by the \textit{small} number $\delta $, is
required to be neither \textit{small }nor \textit{large}.

The relation%
\begin{equation}
\Delta \,\tilde{j}=\left\lfloor O(q\,\delta )\right\rfloor  \label{DeltaTc}
\end{equation}%
implied by this argument indicates that, for any given $q,$ one can always
choose (finitely different) initial data which differ by such a tiny amount
that the corresponding periods are identical, confirming our previous
statement about the \textit{isochronous} character of our model whenever the
parameter $\mu $ is \textit{rational}. But conversely this finding also
implies that, for any set of initial data in the sector under present
consideration (i. e. that characterized by the inequalities $\left\vert \eta
\right\vert <1$ and $\left\vert \bar{\xi}\right\vert +\left\vert
R\right\vert >\left\vert r_{b}\right\vert ,$ and by an additional
specification to identify the physical root $\check{w}(t)$, see \cite{CGSS}%
), if our physical model (\ref{EqMot}) is characterized by an \textit{%
irrational} value of $\mu ,$ see (\ref{mu}), and one replaces this value by
a more and more accurate \textit{rational} approximation of it, see (\ref%
{mupq}) -- as it would for instance be inevitable in any numerical
simulation -- corresponding to larger and larger values of $p$ and
$q,$ then one shall have to choose the two different assignments
of initial data closer and closer to avoid a drastic change of
period -- and for these very close sets of data the motion is
indeed \textit{periodic} with a period (which we are able to
predict, see (\ref{jtilab}) and \cite{CGSS}, but) which becomes
larger and larger the better one approximates the actual,
\textit{irrational} value of $\mu .$ Moreover in any numerical
simulation the accuracy of the computation, in order to get the
correct period, shall also have to increase more and more (with no
limit), because of the occurrence of closer and closer
\textit{near misses} through the time evolution (associated with
the coming into play of \textit{active }branch points sitting on
the circle $B$ closer and closer to the points of intersection
with the circle $\Xi $). And
finally, if one insists in treating the problem with a truly \textit{%
irrational }$\mu ,$ then, no matter how close the initial data are, the
change in the periods becomes \textit{infinite} because the difference $%
\Delta b$ in the number of \textit{active} \textit{square-roots} branch
points on the circle $B$ included \textit{inside} the circle $\Xi $ is
\textit{infinite} (see (\ref{DeltaTa})), signifying that the motion is \textit{%
aperiodic}, and that its evolution is indeed characterized by an \textit{%
infinite} number of \textit{near misses}, making it truly \textit{irregular.}

This phenomenology, together with that of the \textit{near misses }as
described above, illustrates rather clearly the \textit{irregular} character
of the motions of our physical model when the coupling constants have
appropriate values (such as to produce an irrational value of $\mu $ outside
the interval $0<\mu <1$) and the initial data are in the sector identified
above. Note that the Lyapunov coefficients associated with the corresponding
trajectories vanish, because these coefficients -- as usually defined --
compare the difference (after an \textit{infinitely long} time) of two
trajectories that, to begin with, differ \textit{infinitesimally;} whereas
our mechanism causing the \textit{irregular} character of the motion
requires, to come into play, an \textit{arbitrarily small but finite}
difference among the initial data. The difference between these two notions
corresponds to the fact that inside the interval between two \textit{differen%
}t real numbers -- however close they may be -- there always is an \textit{%
infinity} of \textit{rational }numbers; while this is \textit{not}
the case between two real numbers that differ only
\textit{infinitesimally}! This observation suggests that, in an
\textit{applicative} context, the mechanism causing a
\textit{sensitive dependence} on the initial data manifested by
our model may be \textit{phenomenologically} relevant even when no
Lyapunov coefficient, defined in the usual manner, is
\textit{positive}. As already observed previously \cite{CSCFS},
this mechanism is in some sense analogous to that yielding
\textit{aperiodic} trajectories in a triangular billiard with
\textit{irrational} angles; although in that case -- in contrast
to ours -- this outcome is mainly attributable to the essentially
singular character of the corners, and moreover no truly
\textit{irregular} motions emerge.

\bigskip

\section{Outlook}

In this paper we have introduced and discussed a $3$-body problem in the
plane suitable to illustrate a mechanism of transition from \textit{regular}
to \textit{irregular} motions. This model is the simplest one we managed to
manufacture for this purpose. Its simplicity permitted us to discuss in
considerable detail the mathematical structure underlining this
phenomenology: this machinery cannot however be too simple since it must
capture (at least some of) the subtleties associated with the \textit{onset}
of an \textit{irregular} behavior. Therefore in this short paper we were
only able to report our main findings without detailing their proofs, and we
also omitted several other relevant aspects of our treatment (including a
fuller discussion of previous work by others in related areas): this
material shall be presented in a separate, much longer, paper \cite{CGSS},
and probably as well via an electronic version of our findings so as to
supplement their presentation with various animations illustrating these
results and their derivation.

Our main motivation to undertake this research project is the hunch that
this mechanism of transition\textit{\ }have a fairly general validity and be
relevant in interesting applicative contexts. Hence we plan to pursue this
study by focussing on other cases where this mechanism is known to play a
key role, including examples (see, for instance, \cite{CSCFS} and \cite{Ind}%
) featuring a pattern of branch points covering densely an \textit{area} of
the \textit{complex} plane of the independent variable rather than being
confined just to reside densely on a \textit{line} as is the case in the
model treated herein; and eventually to extend the application of this
approach to problems of direct applicative interest.

\begin{figure}[htbp]
\begin{center}
    \noindent\psfig{figure=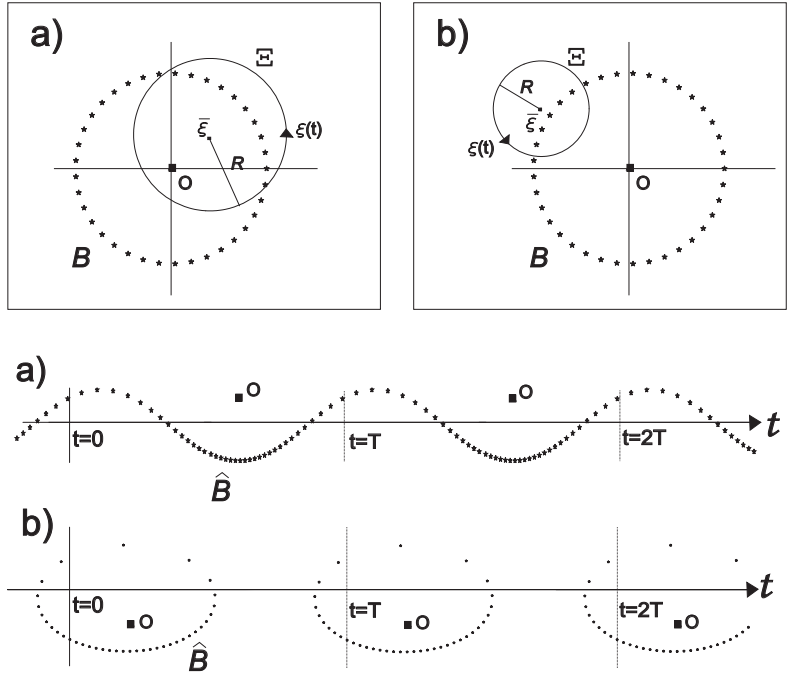,width=4.1in}
    \caption{ Evolutionary paths and branch points of the solutions in the complex $\xi$-plane and $t$-plane for $\mu=15/37$.}
    \label{fig2}
    \end{center}
\end{figure}

In this connection the following final observation is perhaps
relevant. In this paper as well as in others \cite{1,CSCFS,Ind}
 the main focus has been on models
featuring a transition from an \textit{isochronous} to an
\textit{irregular} regime, and in this context much emphasis was
put on the ``trick'' (\ref{zita}) and in particular on the
relationship it entails
between the \textit{periodicity} of the (``physical'') dependent variables $%
z_{n}(t)$ as functions of the \textit{real }independent variable
$t$ (``time'') and the \textit{analyticity} of other, related
(``auxiliary'') dependent variables $\zeta _{n}\left( \tau \right)
$ as functions of a \textit{complex }independent variable $\tau $.
But our findings can also be interpreted \textit{directly} in
terms of the \textit{analytic properties} of the physical
dependent variables $z_{n}(t)$ as functions of the independent
variable $t$ considered itself as a \textit{complex} variable.
Then the time evolution, which corresponded to a uniform travel
round and round on the circle $C$ in the \textit{complex} $\tau
$-plane or equivalently on the circle $\Xi $ in the
\textit{complex} $\xi $-plane, is represented as a uniform travel
to the right along the \textit{real} axis in the \textit{complex
}$t$-plane, while, via the relations (see (\ref{zita}) and
(\ref{KSI}))
\begin{equation}
t=\left( 2\,\rmi\,\omega \right) ^{-1}\,\log \left(
1+2\,\rmi\,\omega \,\tau
\right) =\left( 2\,\rmi\,\omega \right) ^{-1}\,\log \left( \frac{\xi -\bar{\xi}%
}{R}\right) ~,  \label{ttauksi}
\end{equation}%
the pattern of branch points in the \textit{complex} $\tau $-plane or
equivalently in the \textit{complex} $\xi $-plane gets mapped into a
somewhat analogous pattern in the \textit{complex }$t$-plane, \textit{%
repeated periodically} in the \textit{real }direction with period $T,$ see (%
\ref{T}). In particular -- to mention the main features relevant to our
treatment, see above -- the circle $B$ on which the \textit{square-root}
branch points in the \textit{complex} $\xi $-plane sit, gets mapped in the
\textit{complex }$t$-plane into a curve $\hat{B}$ on which sit the \textit{%
square-root} branch points in the \textit{complex} $t$-plane; note that this
curve $\hat{B}$ (in contrast to the circle $B$) \textit{does} now \textit{%
depend} on the initial data. This curve is of course repeated
periodically; it is \textit{closed} and contained in each vertical
slab of width $T$ (see figure \ref{fig2}b) if
the point $\bar{\xi}$ is \textit{outside }$B$ , otherwise it is \textit{open}%
, starting in one slab and ending in the adjoining slab at a point
shifted by the amount $T$; and it does not or does cross (of
course twice in each period) the \textit{real} axis in the
\textit{complex }$t$-plane depending whether, in the
\textit{complex} $\xi $-plane, the two circles $B$ and $\Xi $ do
not or do intersect each other (see figure \ref{fig2}a).
 Likewise, depending whether it is \textit{%
inside} or \textit{outside} the circle $\Xi ,$ the point $\xi =0$ -- which,
as entailed by our analysis, is a highly relevant branch point in the
\textit{complex} $\xi $-plane (unless $0<\mu <1$) -- gets mapped into an
analogous branch point located in each vertical slab \textit{above} or
\textit{below }the \textit{real }axis in the \textit{complex }$t$-plane;
while the other branch point, at $\xi =\infty $ in the \textit{complex} $\xi
$-plane, gets mapped into an analogous branch point located at infinity in
the \textit{lower half} of the \textit{complex }$t$-plane. Clearly the
physical mechanism of \textit{near misses}, which is the main cause of the
eventual \textit{irregularity} of the motion, becomes relevant only for
initial data such that the curve $\hat{B}$ crosses the real axis, thereby
causing (if $\mu $ is \textit{irrational}) an infinity of \textit{%
square-root }branch points of the functions $z_{n}(t)$ to occur \textit{%
arbitrarily close} to the \textit{real} axis in the \textit{complex}%
\thinspace $t$-plane -- branch points which are however \textit{active}
(namely, they actually cause a \textit{near miss} in the physical evolution)
in only some (yet still an infinity) of the infinite number of vertical
slabs in which the \textit{complex} $t$-plane gets now naturally
partitioned. The \textit{near miss} implies that the two particles involved
in it slide past each other from one side or the other depending whether the
corresponding branch point is just above or just below the real axis in the
\textit{complex} $t$-plane. The \textit{sensitive} dependence on the initial
data is due to the fact that any tiny change of them causes some \textit{%
active }branch point in the \textit{complex} $t$-plane which is very close
to real axis to cross over from one side of it to the other, thereby
drastically changing the outcome of the corresponding \textit{near miss}.

This terse discussion shows clearly that the explanation of the \textit{%
irregular} behavior of a dynamical system in terms of travel on a Riemann
surface is by no means restricted to \textit{isochronous} systems. We found
it convenient to illustrate in detail this paradigm by focussing in this
paper on a simple \textit{isochronous} model and by using firstly $\tau $
and then $\xi $ as independent \textit{complex} variables -- but, as
outlined just above, our analysis can also be done -- albeit less neatly --
by using directly the independent \textit{complex} variable $t$; and the
occurrence of a kind of periodic partition of the \textit{complex} $t$-plane
into an infinite sequence of vertical slabs -- characteristic of our \textit{%
isochronous} model -- does not play an essential role to explain the \textit{%
irregular} character of the motion when such a phenomenology does indeed
emerge. The essential point is the possibility to reinterpret the time
evolution as travel on a Riemann surface, the structure of which is
sufficiently complicated to cause an \textit{irregular} motion featuring a
\textit{sensitive dependence} on its initial data. The essential feature
causing such an outcome is the presence of an \textit{infinity} of branch
points \textit{arbitrarily close} to the \textit{real} axis in the \textit{%
complex} $t$-plane, the positions of which, as well as the identification of
which of them are \textit{active}, depends on the \textit{initial data}
nontrivially. The model treated in this paper shows that such a structure
can be complicated enough to cause an \textit{irregular} motion, yet
amenable to a simple mathematical description yielding a rather detailed
understanding of this motion; this suggests the efficacy also in more
general contexts of this paradigm to understand (certain) \textit{irregular}
motions featuring a \textit{sensitive dependence} on their initial data and
possibly even to \textit{predict} their behavior to the extent such a
paradoxical achievement (predicting the unpredictable!) can at all be
feasible.

\bigskip

\ack

It is a pleasure to acknowledge illuminating discussions with Boris
Dubrovin, Yuri Fedorov, Jean-Pierre Fran\c{c}oise, Fran\c{c}ois Leyvraz,
Jaume Llibre, Alexander Mikhailov and Carles Sim\'{o}.

\end{document}